\documentclass[reqno,11pt]{article}
\usepackage{psfig, amsmath, amstexnb, amsthm}
\usepackage{amssymb}  

\def\captionspace{\vspace{2mm}}

\input{stochbif.sty}

\begin{document}


\title{Beyond the Fokker--Planck equation: \\
Pathwise control of noisy bistable systems}
\author{Nils Berglund and Barbara Gentz}
\date{}   

\maketitle

\begin{abstract}
\noindent
We introduce a new method, allowing to describe slowly time-dependent
Langevin equations through the behaviour of individual paths. This
approach yields considerably more information than the computation of
the probability density. The main idea is to show that for
sufficiently small noise intensity and slow time dependence, the vast
majority of paths remain in small space--time sets, typically in the
neighbourhood of potential wells. The size of these sets often has a
power-law dependence on the small parameters, with universal
exponents. The overall probability of exceptional paths is
exponentially small, with an exponent also showing power-law
behaviour.  The results cover time spans up to the maximal Kramers
time of the system.  We apply our method to three phenomena
characteristic for bistable systems: stochastic resonance, dynamical
hysteresis and bifurcation delay, where it yields precise bounds on
transition probabilities, and the distribution of hysteresis areas and
first-exit times. We also discuss the effect of coloured noise. 
\end{abstract}

\leftline{\small{\it Date.\/} October 7, 2001.}
\leftline{\small 2001 {\it PACS numbers.\/} 
02.50.-r, 05.10.Gg, 75.60.-d, 92.40.Cy.
}
\leftline{\small 2000 {\it Mathematical Subject Classification.\/} 
37H20 (primary), 60H10, 34E15, 82C31 (secondary).
}
\noindent{\small{\it Keywords and phrases.\/}
Langevin equation, Fokker--Planck equation, double-well potential,
first-exit time, scaling laws, stochastic resonance, dynamical
hysteresis, bifurcation delay, white noise, coloured noise. 
}  


\section{Introduction}
\label{sec_in}

Noise is often used to model the effect of fast degrees of freedom,
which are too involved to describe otherwise. In statistical physics
and solid state physics, for instance, the influence of a heat bath is
represented by a stochastic Glauber dynamics or a Langevin
equation. In meteorological and climate models, the effect of fast
modes (e.\,g.\ short wavelength modes neglected in a Galerkin
approximation) is often described by noise~\cite{Hasselmann,Arnold2}. 
As a consequence, stochastic differential equations are widely used to
model systems of physical interest, including
ferromagnets~\cite{Martin}, lasers~\cite{Risken,HL,HN},
neurons~\cite{Tu,Longtin}, glacial cycles~\cite{Benzi2}, oceanic
circulation~\cite{Cessi}, biomolecules~\cite{SHD}, and more. 

Among the simpler stochastic models in use is the Langevin equation
with additive white noise 
\begin{equation}
\label{in1}
\6x_t = -\nabla V(x_t,\lambda) \6t + \sigma G(\lambda) \6W_t.
\end{equation}
Here $V$ is a potential, $W_t$ denotes a standard vector-valued Wiener
process (i.\,e., a Brownian motion), and $\sigma$ measures the noise
intensity. For now, we consider $\lambda$ as a fixed parameter, but
below we will be concerned with situations where $\lambda$ varies
slowly in time.  Of course, one may be interested in situations where
noise enters in a different way, for instance $G$ depending on $x$ as
well, or coloured noise. 

There exist different methods to characterize the dynamics of the
Langevin equation \eqref{in1}. A popular approach is to determine the
probability density $p(x,t)$ of $x_t$, which gives all information on
the instantaneous state of the system. For instance, the probability
that $x_t$ belongs to a subset $\cD$ of phase space is given by the
integral of $p(x,t)$ over $\cD$. The density is given by the
normalized solution of the Fokker--Planck equation
\begin{equation}
\label{in2}
\dpar{}t p(x,t) = \nabla \cdot \bigpar{\nabla V(x,\lambda) p(x,t)} +
\nabla \cdot D(\lambda) \nabla p(x,t), 
\end{equation}
where $D = (\sigma^2/2) G G^T$ is the diffusion matrix. In particular, in
the isotropic case $G G^T=\one$, \eqref{in2} admits the stationary
solution
\begin{equation}
\label{in3}
p_0(x) = \frac1N \e^{-2V(x)/\sigma^2},
\end{equation}
where $N$ is the normalization. For small $\sigma$, the stationary
distribution is sharply peaked around the minima of the potential. For
an arbitrary initial distribution, however, the Fokker--Planck
equation cannot be solved in general, and one has to rely on spectral
methods, WKB approximations and the like.

Even if we have obtained a solution of \eqref{in2}, this approach
still has serious shortcomings. The reason is that the probability
density only gives an instantaneous picture of the system. If, for
instance, we want to compute correlation functions such as $\expec{x_s
x_t}$ for $0<s<t$, solving the Fokker--Planck equation with initial
condition $x_0$ is not enough: We need to solve it for {\it all\/}
initial conditions $(x_s,s)$.  Quantities such as the supremum of
$\norm{x_t}$ over some time interval are even harder to handle. 

It is important to take into account that the stochastic differential
equation \eqref{in1} does not only induce a probability distribution
of $x_t$, but also generates a measure on the paths, which contains
much more information. For almost any realization $W_t(\w)$ of the
Brownian motion, and any deterministic initial condition $x_0$, the
solution $\set{x_t(\w)}_{t\geqs0}$ of \eqref{in1} is a continuous
function of time (though not differentiable). The random variable
$x_t(\w)$ for fixed $t$ is only one of many interesting random
quantities which can be associated with the stochastic process. 

First-exit times form an important class of such alternative random
variables, and have been studied in detail. If $\cD$ is a (measurable)
subset of phase space, the first-exit time of $x_t$ from $\cD$ is
defined as 
\begin{equation}
\label{in4}
\tau_\cD \defby \inf \bigsetsuch{t>0}{x_t\not\in\cD}.
\end{equation}
For instance, if $\cD$ is a set of the form $\setsuch{x}{V(x)\leqs
V_1}$, containing a unique equilibrium point which is stable, 
then the distribution of $\tau_\cD$ is asymptotically
exponential, with expectation behaving in the small-noise limit like
Kramers' time 
\begin{equation}
\label{in5}
T_{\math{Kramers}} = \e^{2(V_1-V_0)/\sigma^2},
\qquad\text{where\ } V_0 \defby \min_{x\in\cD} V(x).
\end{equation} 
A mathematical theory allowing to estimate first-exit times for general
$n$-dimensional systems (with a drift term not necessarily
deriving from a potential) has been developed by Freidlin and
Wentzell~\cite{FW}. In specific situations, more precise results are
available, for instance the following. 
If $\cD$ contains a unique, stable equilibrium
point, subexponential corrections to the asymptotic
expression~\eqref{in5} are known, even for possibly time-dependent
drift terms~\cite{Azencott,FJ}. The case where $\cD$ 
contains a saddle as unique equilibrium point has been considered by
Kifer in the seminal paper~\cite{Kifer}. The situation where $\cD$
contains a stable equilibrium in its interior and a saddle on its
boundary is dealt with 
in~\cite{MS}, employing the method of matched asymptotic expansions. 
The limiting behaviour of the distribution of the first-exit time from
a neighbourhood of a unique stable equilibrium point as well 
as from a neighbourhood of a saddle point has been obtained by
Day~\cite{Day1, Day2}.

We remark that more recently, another approach has been introduced,
which mimics concepts from the theory of dissipative dynamical systems
\cite{CF1,Schmalfuss,Arnold}. The main idea is that for a given
realization of the noise (i.\,e., in a quenched picture), paths with
different initial conditions may converge to an attractor, which has
similar properties as deterministic attractors. Of course, in
experiments, this random attractor is only visible if we manage to
repeat the experiment many times with the same realization of noise. 

The method of choice to study the Langevin equation \eqref{in1} also
depends on the time scale we are interested in. Consider for instance a
one-dimensional double-well potential $V(x)=-\frac12ax^2+\frac14bx^4$, $a,
b>0$, with a barrier of height $H=a^2/4b$. Assume that $x_0$ is
concentrated at the bottom $x=\smash{\sqrt{a/b}}$ of the right-hand
potential well. If the noise is sufficiently weak, paths are likely to
stay in the right-hand well for a long time. The distribution of $x_t$ will
first approach a Gaussian in a time of order 
\begin{equation}
\label{in6}
T_{\math{relax}} = \frac1c, 
\end{equation}
where $c=2a$ is the curvature at the bottom of the well (the variance of
the Gaussian is approximately $\sigma^2/(2c)$). With overwhelming
probability, paths will remain inside the same potential well, for all
times significantly shorter than Kramers' time $T_{\math{Kramers}} =
\e^{2H/\sigma^2}$. Only on longer time scales will the density of $x_t$
approach the bimodal stationary density \eqref{in3}. The dynamics will thus
be very different on the time scales $t\ll T_{\math{relax}}$,
$T_{\math{relax}}\ll t\ll T_{\math{Kramers}}$, and $t\gg
T_{\math{Kramers}}$. Random attractors can be reached only in the last
regime. In particular, results in \cite{CF2} stating that for the
double-well potential $V$, the random attractor almost surely consists of
one (random) point apply to the regime $t\gg T_{\math{Kramers}}$. 

In the present work, we are interested in situations where the parameter
$\lambda$ varies slowly in time, that is, we consider stochastic
differential equations (SDEs) of the form 
\begin{equation}
\label{in7}
\6x_t = -\nabla V(x_t,\lambda(\eps t)) \6t + \sigma G(\lambda(\eps t)) \6W_t.
\end{equation}
Such situations occur if the system under consideration is slowly forced,
for instance by an external magnetic field, by climatic changes, or by a
varying energy supply. Note that  the probability density of $x_t$ still
obeys a Fokker--Planck equation, but there will be no stationary solution
in general. 

The main idea of our approach to equations of the form \eqref{in7} is
the following. For sufficiently small $\eps$ and $\sigma$, and on an
appropriate time scale, we show that the paths $\set{x_t}_{t\geqs0}$
can be divided into two classes. The first class consists of those
paths which remain in certain space--time sets, typically in the \nbh\
of potential wells (but in some cases, they may switch potential
wells). The geometry and size of these sets depends on noise intensity
and shape of potential.  The second class consists of the remaining
paths, which we do not try to describe in detail, but whose overall
probability is small, typically exponentially small in some
combination of $\eps$ and $\sigma$. In this way, we obtain in
particular concentration results on the density of $x_t$ without
solving the Fokker--Planck equation, but we also obtain global
information on the stochastic process, including correlations,
first-exit times, transition probabilities and, for instance, the
shape of hysteresis cycles. 

The slow time dependence of $\lambda$ introduces a new time scale. If, for
instance, $\lambda$ is periodic with period $1$, then \eqref{in7} depends
periodically on time with period $T_{\math{forcing}}=1/\eps$. Since the
shape of the potential changes in time, the definitions \eqref{in5} and
\eqref{in6} of the Kramers and relaxation times no longer make sense. We
can, however, define these time scales by 
\begin{equation}
\label{in8}
T_{\math{Kramers}}^{(\math{max})} = \e^{2H_{\math{max}}/\sigma^2}
\qquad\qquad
\text{and}
\qquad\qquad
T_{\math{relax}}^{(\math{min})} = \frac1{c_{\math{max}}},
\end{equation}
where $H_{\math{max}}$ and $c_{\math{max}}$ denote, respectively, the
maximal values of barrier height and curvature of a potential well over one
period. Our results typically apply to the regime 
\begin{equation}
\label{in9}
T_{\math{relax}}^{(\math{min})} \ll T_{\math{forcing}} \ll
T_{\math{Kramers}}^{(\math{max})},
\end{equation}
that is, we require that $\eps\ll c_{\math{max}}$ and $\sigma^2\ll
2H_{\math{max}}/\abs{\log\eps}$. The minimal curvature and barrier height,
however, are allowed to become small, or even to vanish.

We can describe the paths' behaviour on time intervals including many 
periods of the forcing, as long as they are significantly shorter than the
maximal Kramers time. We find it convenient to measure time in units of the
forcing period. After scaling $t$ by a factor $\eps$, the relevant time
scales become
\begin{equation}
\label{in10}
T_{\math{relax}}^{(\math{min})} = \frac\eps{c_{\math{max}}},
\qquad
T_{\math{forcing}} = 1,
\qquad
T_{\math{Kramers}}^{(\math{max})} = \eps\e^{2H_{\math{max}}/\sigma^2}.
\end{equation}
When scaling the Brownian motion, we should keep in mind its diffusive
nature, which implies that in the new units, its standard deviation grows
like $\smash{\sqrt{t/\eps}}$. Equation~\eqref{in7} thus becomes
\begin{equation}
\label{in11}
\6x_t = -\frac1\eps \nabla V(x_t,\lambda(t)) \6t 
+ \frac\sigma{\sqrt\eps} \, G(\lambda(t)) \6W_t.
\end{equation}
In the deterministic case $\sigma=0$, we will sometimes write this equation
in the form $\eps\dot x=-\nabla V(x,\lambda(t))$, which is customary in
singular perturbation theory. Throughout the paper, we will assume that $V$,
$G$ and $\lambda$ are smooth functions, that $V(x,\lambda)$ grows at least
like $\norm{x}^2$ for large $\norm{x}$, and that the matrix elements of $G$,
as well as $\lambda$, have uniformly bounded absolute values. A large part
of the paper is devoted to one-dimensional systems, and we come back to the
multidimensional case in Section~\ref{sec_g}. 

\begin{figure}
 \centerline{\psfig{figure=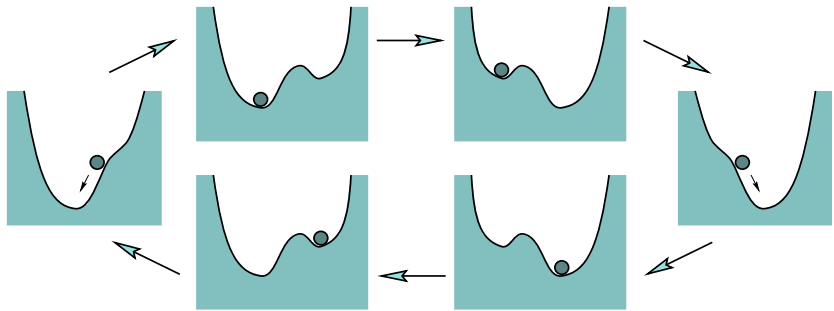,width=110mm,clip=t}}
 \captionspace
 \caption[]
 {Example of a periodically forced double-well potential. In the limit of
 infinitely slow forcing, an overdamped particle tracks the bottom of
 potential wells. However, its position is not determined by the
 instantaneous value of the forcing alone. The system describes a
 hysteresis cycle, whose shape depends on the frequency of the forcing.
 Noise may kick the particle over the potential barrier, and thus
 influences the shape of the hysteresis cycle.}
\label{fig_potential}
\end{figure}

The aim of this paper is to illustrate our methods by applying them to a
number of physically interesting examples. We thus emphasize the conceptual
aspects, and refrain from giving mathematical details of the proofs, which
will be presented elsewhere \cite{BG1,BG2,BG3}.  

We start, in Section~\ref{sec_sw}, by explaining the basic ideas in the
simplest situation, where most paths remain concentrated near the bottom of
a potential well. We also briefly describe the dynamics near a saddle. The
three subsequent sections are devoted to more interesting cases in which
paths may jump between the wells of a double-well potential. Note that we
restrict our attention to double-well potentials only to keep the
presentation simple, but the dynamics in more complicated multi-well
potentials can be described by the same approach. 

In Section~\ref{sec_sr}, we discuss the phenomenon of stochastic resonance,
where noise allows transitions between potential wells which would be
impossible in the deterministic case. We compute the threshold noise
intensity needed for transitions to be likely, and show that most paths are
close, in a natural geometrical sense, to a periodic function. This
provides an alternative quantitative measure of the signal's periodicity to
the commonly used signal-to-noise ratio. 	

Section~\ref{sec_hy} is devoted to hysteresis, which is also characteristic
for forced bistable systems. For sufficiently strong forcing, one of the two
potential wells disappears, causing trajectories to switch between
potential wells even when no noise is present (\figref{fig_potential}). As
a result, even in the adiabatic limit, the instantaneous value of the
parameter $\lambda$ does not suffice to determine the state of the system:
Solutions follow hysteresis cycles. In absence of noise, their area is
known to scale in a nontrivial way with the frequency of the forcing. Our
methods allow us to characterize the distribution of the random hysteresis
area for positive noise intensity. In particular, we show that for noise
intensities above an amplitude-depending threshold, the typical area no
longer depends, to leading order, on amplitude or frequency of the
forcing. 

In Section~\ref{sec_bd}, we consider the effect of additive noise on
systems with spontaneous symmetry breaking, i.\,e., when the potential
transforms from single to double well. In the deterministic case, solutions
are known to track the saddle for a considerable time before falling into
one of the potential wells, a phenomenon known as bifurcation delay. We
characterize the effect of additive noise on this delay, and on the
probability to choose one or the other potential well after the
bifurcation. Furthermore, we give results on the concentration of paths near
potential wells which allow, in particular, to determine the optimal
relation between speed of parameter drift and noise intensity for an
experimental determination of the bifurcation diagram. 

Finally, Section~\ref{sec_g} contains some generalizations. We first discuss
analogous results to those of Section~\ref{sec_sw} for multidimensional
potentials. This formalism allows us to treat the effect of the simplest
kind of coloured noise, given by an Ornstein--Uhlenbeck process,
in a natural way. We conclude by discussing the dependence of
previously discussed phenomena on noise colour.

\subsubsection*{Acknowledgements:}  
We are grateful to Anton Bovier for helpful comments on a preliminary
version of the manuscript. N.\,B. thanks the WIAS for kind hospitality.

\goodbreak


\section{Near wells and saddles}
\label{sec_sw}

We start by discussing situations in which the noise intensity is
sufficiently small, compared to the depth of a given potential well, for
paths to remain concentrated near the bottom of the well during a long time
interval. In Section~\ref{ssec_swl}, the solvable linear case is used to
compare the information provided by the Fokker--Planck equation and by the
pathwise approach. In Section~\ref{ssec_swn}, we show that our method
naturally extends to the nonlinear case. We briefly describe the dynamics
near a saddle in Section~\ref{ssec_es}. 


\subsection{Linear case}
\label{ssec_swl}

It is instructive to consider first the case of a linear force (that is, of
a quadratic potential), which can be solved completely. A general
one-dimensional, time-dependent quadratic potential can be written as 
\begin{equation}
\label{swl1}
V(x,t) = \frac12 c(t) \bigpar{x-x^\star(t)}^2,
\end{equation}
where $x^\star(t)$ is the location of the potential minimum, and $c(t)$ is
the curvature of the potential. The SDE~\eqref{in7} takes the form 
\begin{equation}
\label{swl2}
\6x_t = \frac1\eps a^\star(t) \bigpar{x_t-x^\star(t)} \6t 
+ \frac\sigma{\sqrt\eps} g(t) \6W_t,
\end{equation}
where $a^\star(t) = -c(t)$. Throughout this paper, we will use $a^\star$ to
denote the linearization of the force at an equilibrium point $x^\star$,
with $a^\star<0$ if the equilibrium is stable, and $a^\star>0$ if it is
unstable. In this subsection and the following one, we consider the stable
case, and assume that $a^\star(t)\leqs -a_0$ for all $t$ under
consideration, where $a_0$ is a positive constant. For simplicity, we shall
assume that the functions $x^\star(t)$ and $a^\star(t)$, as well as
$g(t)$, are real-analytic. 

Let us first investigate the probability density $p(x,t)$ of $x_t$. The
Fokker--Planck equation being linear, it can be easily solved. Assume for
simplicity that the distribution of $x_0$ is Gaussian, with expectation
$\expec{x_0}$ and (possibly zero) variance $\variance\set{x_0}$. (In
addition, we always assume the initial distribution to be independent of the
Brownian motion.) Then $x_t$ has a Gaussian distribution for any $t>0$, with
density
\begin{equation}
\label{swl3:new}
p(x,t) = \frac1{\sqrt{2\pi\variance\set{x_t}}}
\exp\biggset{-\frac{(x-\expec{x_t})^2}{2\variance\set{x_t}}},
\end{equation}
where expectation and variance of $x_t$ obey the ODEs
\begin{align}
\label{swl4:new}
\dtot{}t \expec{x_t} &= \frac1\eps a^\star(t) \bigpar{\expec{x_t}-x^\star(t)}
\\
\label{swl5:new}
\dtot{}t \variance\set{x_t} &= \frac2\eps a^\star(t) \variance\set{x_t} +
\frac{\sigma^2}\eps g(t)^2.
\end{align}
Note that $\expec{x_t}$ coincides with the deterministic solution $\xdet_t$
of Equation~\eqref{swl2} for $\sigma=0$, with initial condition $\xdet_0 =
\expec{x_0}$: 
\begin{equation}
\label{swl4}
\expec{x_t} = \xdet_t = \xdet_0 \e^{\alpha^\star(t)/\eps}  
- \frac1\eps \int_0^t \e^{\alpha^\star(t,s)/\eps} a^\star(s) x^\star(s) \6s,
\end{equation}
where we use the notations
\begin{equation}
\label{swl5}
\alpha^\star(t,s) = \int_s^t a^\star(u) \6u, \qquad
\alpha^\star(t) = \alpha^\star(t,0).
\end{equation}
Our stability assumption $a^\star(t)\leqs -a_0$ $\forall t$ implies that
$\alpha^\star(t,s) \leqs -a_0(t-s)$ for $t>s$. Hence the first term on the
right-hand side of \eqref{swl4} decreases exponentially fast: It is at most
of order $\eps$ after time $\eps\abs{\log\eps}/a_0$,  at most of order
$\eps^2$ after time $2\eps\abs{\log\eps}/a_0$, and so on.  

We expect $\xdet_t$ to follow adiabatically the slowly drifting bottom of
the potential well. To make this apparent, we evaluate the second term on
the right-hand side of \eqref{swl4} by integration by parts: 
\begin{equation}
\label{swl6}
- \frac1\eps \int_0^t \e^{\alpha^\star(t,s)/\eps} a^\star(s) x^\star(s) \6s
= x^\star(t) - x^\star(0)\e^{\alpha^\star(t)/\eps} - 
\int_0^t \e^{\alpha^\star(t,s)/\eps} \dot x^\star(s) \6s.
\end{equation}
By successive integrations by parts, we find that the general solution of
\eqref{swl4:new} can be written as 
\begin{equation}
\label{swl7}
\expec{x_t} = \xdet_t = 
\xbdet_t + (\xdet_0 - \xbdet_0) \e^{\alpha^\star(t)/\eps},
\end{equation}
where $\xbdet_t$ is a particular solution of \eqref{swl4:new}, admitting the
asymptotic expansion\footnote{The asymptotic series does not converge in
general, but it admits expansions to any order in $\eps$, with a remainder
which can be controlled.} 
\begin{equation}
\label{swl8}
\xbdet_t = x^\star(t) + \eps \frac{\dot x^\star(t)}{a^\star(t)} + 
\eps^2 \frac1{a^\star(t)} \dtot{}t \biggpar{\frac{\dot
x^\star(t)}{a^\star(t)}} + \dotsb
\end{equation}
Since $a^\star(t)$ is negative, $\xbdet_t$ tracks the bottom $x^\star(t)$
of the potential well with a small lag: $\xbdet_t < x^\star(t)$ if
$x^\star(t)$ moves to the right, and $\xbdet_t > x^\star(t)$ if
$x^\star(t)$ moves to the left. The particular solution \eqref{swl8} is
called \defwd{adiabatic solution} or \defwd{slow solution}.
Relation~\eqref{swl7} expresses the fact that all solutions of
\eqref{swl4:new} are attracted exponentially fast by the adiabatic solution
$\xbdet_t$. 

The variance of $x_t$ can be computed in a similar way. The solution of
\eqref{swl5:new} is given by 
\begin{equation}
\label{swl11}
\variance\set{x_t} = \variance\set{x_0} \e^{2\alpha^\star(t)/\eps} + 
\frac{\sigma^2}{\eps} \int_0^t \e^{2\alpha^\star(t,s)/\eps} g(s)^2 \6s.
\end{equation}
The behaviour of the variance is very similar to the behaviour of the
deterministic solution \eqref{swl4}: The initial condition
$\variance\set{x_0}$ is forgotten exponentially fast, and in analogy with
\eqref{swl7} and \eqref{swl8}, we can write 
\begin{equation}
\label{swl12}
\variance\set{x_t} = \bv(t) +  
\bigpar{\variance\set{x_0} - \bv(0)} \e^{2\alpha^\star(t)/\eps},
\end{equation}
where $\bv(t)$ is a particular solution of \eqref{swl5:new}, which admits the
asymptotic expansion
\begin{equation}
\label{swl13}
\bv(t) = \frac{\sigma^2}{2\abs{a^\star(t)}} \biggbrak{g(t)^2 + \eps \dtot{}t
\biggpar{\frac{g(t)^2}{2a^\star(t)}} + \dotsb}.
\end{equation}

The fact that $x_t$ has a Gaussian distribution implies in particular that
for any $t>0$, 
\begin{equation}
\label{swl14}
\Bigprob{\abs{x_t-\xdet_t}>h \sqrt{\variance\set{y_t}}\mskip1.5mu} = 2 
\int_{h\sqrt{\variance\set{y_t}}}^\infty p(\xdet_t+y,t) \6y  
\leqs \e^{-h^2/2}.
\end{equation}
Hence, the distribution of $x_t$ is concentrated in an interval of width
$\sqrt{2\variance\set{y_t}}$ around $\xdet_t$, which behaves asymptotically
like $\sigma g(t)/\sqrt{\abs{a^\star(t)}}$ by \eqref{swl12} and
\eqref{swl13}. In words, the spreading of $x_t$ is proportional to the
noise intensity and inversely proportional to the square root of the
curvature of the potential: Flatter potentials give rise to a larger
spreading of the distribution. 

Up to now, we have only studied the probability density of $x_t$. However,
even if the density is concentrated near the bottom of the well at all
times, this does not exclude that the path $\set{x_s}_{0\leqs s\leqs t}$
makes occasional excursions away from $x^\star$. From now on, we consider
the initial condition $x_0$ as deterministic, so that the path depends only
on the realization of the Brownian motion. The solution of the
SDE~\eqref{swl2} can be written as
\begin{equation}
\label{swl10}
x_t = \xdet_t + y_t, 
\qquad\qquad
y_t =  \frac{\sigma}{\sqrt\eps} 
\int_0^t \e^{\alpha^\star(t,s)/\eps} g(s)\6W_s. 
\end{equation}
The process $y_t$ is a generalization of an Ornstein--Uhlenbeck process,
with time-dependent damping and diffusion. Ideally, we would like to
estimate the probability that the path leaves a strip of (time-dependent)
width proportional to $\sqrt{\variance\set{x_t}}$, and centred at
$\xdet_t$. This turns out to be difficult because the variance may
change quickly for $t$ very close to $0$, due to the first term on the
right-hand side of \eqref{swl11}. To avoid these technical
complications, we use a strip of width proportional to
$\sqrt{\bv(t)}$, defined by  
\begin{equation}
\label{swl15}
\cB(h) = \bigsetsuch{(x,t)}{\abs{x-\xdet_t} < h\sqrt{\bv(t)}}. 
\end{equation}
Note that $\cB(h)$ coincides, up to order $\eps$, with the set of points
where $V(x,t)$ is smaller than $V(\xdet_t,t)+(\frac12 h\sigma g(t))^2$. 
Showing that the path $\set{x_s}_{0\leqs s\leqs t}$ is likely to remain in
$\cB(h)$ is equivalent to showing that the \defwd{first-exit time} of $x_s$
from $\cB(h)$, defined by 
\begin{equation}
\label{swl16}
\tau_{\cB(h)} = \inf \bigsetsuch{s\geqs 0}{(x_s,s)\not\in\cB(h)},
\end{equation}
is unlikely to be smaller than $t$. In fact, the following
probabilities are equivalent (the superscripts refer to the initial
condition): 
\begin{align}
\nonumber
\bigprobin{0,x_0}{\tau_{\cB(h)}<t} 
&= \bigprobin{0,x_0}{\exists s\in[0,t) \colon (x_s,s) \not\in \cB(h)} \\
\label{swl17}
&= \biggprobin{0,x_0}{\sup_{0\leqs s < t}
\frac{\abs{x_s-\xdet_s}}{\sqrt{\bv(s)}} \geqs h}.
\end{align}
The following result is a straightforward consequence of standard
exponential bounds on the supremum of stochastic integrals, extended to
integrals as appearing in \eqref{swl10}. The proof, given in
\cite[Proposition~3.4]{BG1} for constant $g$, also applies
here.\footnote{The generalization of the proof to $g$ bounded away from zero 
is trivial, but the result also holds if $g$ vanishes. In fact, 
a sufficient condition is that $\bv(t)/\bv(s) = 1 + \Order{\eps}$ 
whenever $t-s=\Order{\eps^2}$, which can be checked using the 
asymptotic expansion~\eqref{swl13}.}

\begin{figure}
 \centerline{\psfig{figure=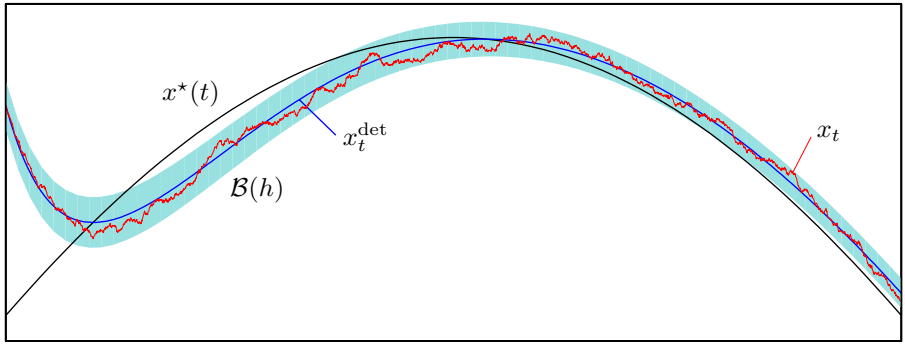,width=120mm,clip=t}}
 \captionspace
 \caption[]
 {A sample path $x_t$ of the linear equation~\eqref{swl2} for
 $a^\star(t)=-4+2\sin(4\pi t)$, $x^\star(t)=\sin(2\pi t)$ and
 $g(t)\equiv1$. Parameter values are $\eps=0.04$ and $\sigma=0.025$. The
 shaded region is the set $\cB(h)$ for $h=3$, centred at the deterministic
 solution $\xdet_t$ starting at the same point as $x_t$. The width of
 $\cB(h)$ is of order $h\sigma/\smash{\sqrt{a^\star(t)}}$.
 Proposition~\ref{prop_swl} states that the probability of $x_t$ leaving
 $\cB(h)$ before times of order $1$ decays roughly like
 $\smash{\e^{-h^2/2}}$.}
\label{fig_stable}
\end{figure}

\begin{prop}
\label{prop_swl}
For all $t$ and $h>0$, 
\begin{equation}
\label{swl18}
\bigprobin{0,x_0}{\tau_{\cB(h)}<t} 
\leqs C(t,\eps) \e^{-\kappa h^2},
\end{equation}
where 
\begin{equation}
\label{swl19}
C(t,\eps) = \frac{\abs{\alpha^\star(t)}}{\eps^2} + 2 
\qquad\qquad
\text{and}
\qquad\qquad
\kappa = \frac12 - \Order{\eps}.
\end{equation}
\end{prop}

Note that only the prefactor $C(t,\eps)$ is time-dependent. The exponential
factor $\e^{-\kappa h^2}$ is small as soon as $h\gg 1$, so that the paths
$\set{x_s}_{0\leqs s\leqs t}$  are concentrated in a neighbourhood of order
$\smash{\sqrt{\bv(s)}}$ of the deterministic solution up to time $t$ (see
\figref{fig_stable}). More precisely, paths are unlikely to leave the strip
$\cB(h)$ of width $h\smash{\sqrt{\bv}}$ before time $t$ provided
$h^2\gg\log C(t,\eps)$. 

The bound \eqref{swl18} is useful for times significantly shorter than
Kramers' time, which is of order $\eps\e^{2H/\sigma^2}$ (to reach points
where the potential has value $H$). For longer times, the prefactor
$C(t,\eps)$ becomes sufficiently large to counteract the term $\e^{-\kappa
h^2}$ for any reasonable $h$, which is natural, as we cannot expect paths
to remain concentrated near $\xdet_t$ on such long time scales. On
polynomial time scales of order $\sigma^{-k}$, however, large excursions
are very unlikely. 

The estimate \eqref{swl18} has been designed to yield an optimal exponent
for noise intensities scaling like a power of $\eps$. We do not expect the
prefactor $C(t,\eps)$ to be optimal, but for times and noise intensities
polynomial in $\eps$, it leads to subexponential corrections. However,
if we do not care for the precise exponent, \eqref{swl19} can be
replaced by  
\begin{equation}
\label{swl20}
C(t,\eps) = \frac{\abs{\alpha^\star(t)}}{\eps} + 2 
\qquad\qquad
\text{and}
\qquad\qquad
\kappa > 0.
\end{equation}
The denominator $\eps$ in $C$ is due to the fact that we work in slow time. 


\subsection{Nonlinear case}
\label{ssec_swn}

We consider now the motion in a more general, nonlinear potential of the form
\begin{equation}
\label{swn1}
V(x,t) = \frac12 c(t) \bigpar{x-x^\star(t)}^2 +
\bigOrder{\bigpar{x-x^\star(t)}^3},
\end{equation}
admitting a local minimum at $x^\star(t)$. As before, we assume that the
curvature $c(t)$ is bounded below by a positive constant for all times $t$.
We do not exclude, however, that $V$ has other potential wells than the one
at $x^\star(t)$. 

The probability density can no longer be computed exactly in general,
although it seems plausible that a distribution initially concentrated near
$x^\star(0)$ will remain concentrated near $x^\star(t)$, on a certain time
scale. In fact, Proposition~\ref{prop_swl} naturally extends to the
nonlinear case. 

Consider first the deterministic case $\sigma=0$. A result due to Grad\v
ste\u\i n and Tihonov \cite{Grad,Tihonov}, which is related to the
adiabatic theorem of quantum mechanics, states that 
\begin{itemiz}
\item	there exists a particular solution $\xbdet_t$ of the deterministic
equation $\eps\dot x = -\sdpar Vx(x,t)$ tracking the bottom of the potential
well at a distance of order $\eps$;
\item	any solution $\xdet_t$ starting in a \nbh\ of $x^\star(0)$ (in fact,
inside the potential well) approaches $\xbdet_t$ exponentially fast in
$t/\eps$.  
\end{itemiz}
Let us fix a deterministic initial condition $x_0=\xdet_0$ such that
$\xdet_t$ is attracted by $\xbdet_t$. We introduce the notations 
\begin{equation}
\label{swn2}
a(t) = -\dpar{^2V}{x^2} (\xdet_t,t), \qquad
\alpha(t,s) = \int_s^t a(u) \6u \quad
\text{and}\quad
\alpha(t) = \alpha(t,0)
\end{equation}
for the curvature of the potential at $\xdet_t$ and the analogue quantities
to \eqref{swl5}. Note that Tihonov's result implies that $a(t)$ 
asymptotically approaches $-c(t)+\Order{\eps}$. The difference
$y_t=x_t-\xdet_t$ satisfies an SDE of the form
\begin{equation}
\label{swn2b}
\6y_t = \frac1\eps \bigbrak{a(t)y_t + b(y_t,t)}\6t +
\frac\sigma{\sqrt\eps}g(t)\6W_t,
\end{equation}
where $b(y,t)=\Order{y^2}$ describes the effect of nonlinearity. 
We define again a strip $\cB(h)$ as in \eqref{swl15}, with 
\begin{equation}
\label{swn3}
\bv(t) = \bv(0)\e^{2\alpha(t)/\eps} + \frac{\sigma^2}\eps \int_0^t
\e^{2\alpha(t,s)/\eps} g(s)^2 \6s. 
\end{equation}
Our results work for any $\bv(0)$ larger than a positive constant
independent of $\eps$. A convenient choice is $\bv(0)=\sigma^2
g(0)^2/(2\abs{a^\star(0)})$: Then the fact that $\xdet_t$ approaches
exponentially fast a \nbh\ of order $\eps$ of $x^\star(t)$ implies that 
\begin{equation}
\label{swn3b}
\bv(t) = \sigma^2  \biggbrak{\frac{g(t)^2}{2\abs{a^\star(t)}} + \Order{\eps} +
\Order{\abs{x_0-x^\star(0)}\e^{-\text{\it const }t/\eps}}}.
\end{equation}
Proposition~\ref{prop_swl} generalizes to 

\begin{theorem}
\label{thm_swn}
There exists a constant $h_0$, independent of $\sigma$ and $\eps$, such that
for all $h\leqs h_0/\sigma$, 
\begin{equation}
\label{swn4}
\bigprobin{0,x_0}{\tau_{\cB(h)}<t} 
\leqs C(t,\eps) \e^{-\kappa h^2},
\end{equation}
where 
\begin{equation}
\label{swn5}
C(t,\eps) = \frac{\abs{\alpha(t)}}{\eps^2} + 2 
\qquad\qquad
\text{and}
\qquad\qquad
\kappa = \frac12 - \Order{\eps} - \Order{\sigma h}.
\end{equation}
\end{theorem}

The interpretation is the same as in the linear case: Paths are
concentrated, for times significantly shorter than Kramers' time, in a
strip of width proportional to $\smash{\sqrt{\bv(t)}}$ around
$\xdet_t$.

The proof is identical to the one of \cite[Theorem~2.4]{BG1}. The
main idea is to show that if the solution of the equation
linearized around $\xdet_t$ remains in a strip $\cB(h)$, then the solution
of the nonlinear equation \eqref{swn2b} almost surely remains in the
slightly larger strip $\cB(h[1+\Order{\sigma h}])$. 

The main difference between the nonlinear and the linear case is the
condition $h\leqs h_0/\sigma$, which stems from the requirement that the
linear term $a(t)y_t$ in Equation~\eqref{swn2b} should dominate the
nonlinear term $b(y_t,t)$ for all $(x_t,t)\in\cB(h)$. Because of this
condition, the result \eqref{swn4} is useful for $\sigma^2\ll \kappa
h_0^2/\log C(t,\eps)$. It is, however, possible to derive bounds for larger
deviations under additional assumptions on the potential:

\begin{prop}
\label{prop_swn}
Assume that there are constants $L_0>0$, $K>0$ and $n\geqs2$ such that 
\begin{equation}
\label{swn6}
x \dpar Vx(x,t) \geqs K \abs{x}^n 
\end{equation}
whenever $\abs{x}\geqs L_0$ and $t\geqs 0$. Then there exist constants $C,
\kappa > 0$ such that 
\begin{equation}
\label{swn7}
\biggprobin{0,x_0}{\sup_{0\leqs s\leqs t} \abs{x_s} \geqs L} 
\leqs C \Bigpar{\frac t\eps + 1} \e^{-\kappa L^n/\sigma^2}
\end{equation}
for all $t\geqs0$, $L\geqs L_0$ and $\abs{x_0}\leqs L_0/2$.
\end{prop}

This result is a generalization of \cite[Proposition~4.3]{BG3}, where the
case $n=4$ was treated. 

We remark in passing that the bounds \eqref{swn4} and \eqref{swn6} are
sufficient to provide estimates on the moments of the distribution of
$x_t$, without solving the Fokker--Planck equation 
(c.\,f.~\cite[Corollary~4.6]{BG3}):

\begin{cor}
\label{cor_swn}
Assume that \eqref{swn6} holds for some $n\geqs2$. Then 
\begin{equation}
\label{swn8}
\bigexpecin{0,x_0}{\abs{x_t - \xdet_t}^{2k}} \leqs (2k-1)!! M^k \bv(t)^k
\end{equation}
for some constant $M$, all integers $k$ and all $t\geqs0$, provided 
$\sigma \leqs c_0/\log(1+t/\eps)$ for a sufficiently small constant $c_0$.
\end{cor}

Note that the Cauchy--Schwarz inequality immediately implies bounds on odd
moments as well. The bounds on the moments are those of a Gaussian
distribution with variance $M\bv(t)$, even if the potential $V$ has
multiple wells. The reason is that on the time scale under consideration,
solutions of the SDE do not have enough time to cross a potential barrier
and reach another potential well. 


\subsection{Escape from a saddle}
\label{ssec_es}

Assume now that the potential $V(x,t)$ admits a saddle at $x^\star(t)$ for
all times under consideration. In the deterministic case, a particular
solution $\xhatdet_t$ is known to track the saddle at a distance of order
$\eps$, separating the basins of attraction of two neighbouring potential
wells. Trajectories starting near $\xhatdet_t$ will depart from it
exponentially fast, but if the initial separation $\abs{x_0-\xhatdet_0}$ is
exponentially small, the time required to reach a distance of order one
from the saddle may be quite long. 

Noise will help kicking $x_t$ away from $\xhatdet_t$, and thus reduce the
time necessary to leave a \nbh\ of the saddle. In order to describe this
effect, we consider the deviation $y_t=x_t-\xhatdet_t$, which satisfies the
SDE
\begin{equation}
\label{es1}
\6y_t = \frac1\eps \bigbrak{\ha(t)y_t + \hb(y_t,t)}\6t +
\frac\sigma{\sqrt\eps}g(t)\6W_t,
\end{equation} 
where $\ha(t)\geqs a_0>0$ is the curvature of the potential at
$\xhatdet_t$, and $\hb(y,t)=\Order{y^2}$. We now describe the dynamics in a
small \nbh\ of the adiabatic solution tracking the saddle, where diffusion
prevails over drift\footnote{A possible way to compare the influence of
drift and noise is by It\^o's formula, which yields, for $\hb=0$,  
$\6\,(y_t^2) = (1/\eps) \brak{2\ha(t)y_t^2 + \sigma^2g(t)^2}\6t +
(2\sigma/\sqrt\eps)g(t)y_t\6W_t$. 
The expression in brackets is dominated by the noise term $\sigma^2 g(t)^2$
for all $(x_t,t)$ in $\cB(1)$, while the deterministic term $2\ha(t)y_t^2$
prevails outside this domain.}, defined by
\begin{equation}
\label{es2}
\cB(h) = \Bigsetsuch{(x,t)}{\abs{x-\xhatdet_t} < 
\frac{h \sigma g(t)}{\sqrt{2\ha(t)}}}.
\end{equation}
The following result, which is proved in the same way as
\cite[Proposition~3.10]{BG1}, shows that the first-exit time
$\tau_{\cB(h)}$ of $x_t$ from $\cB(h)$ is likely to be small.

\begin{theorem}
\label{thm_es}
Assume that $g$ is bounded below by $L\eps$ for a sufficiently large
constant $L$. Then for all $h\leqs1$ and all initial conditions
$(x_0,0)\in\cB(h)$,
\begin{equation}
\label{es3}
\bigprobin{0,x_0}{\tau_{\cB(h)}>t} \leqs C \exp\Bigset{-
\frac{\kappa}{h^2} \frac1\eps \int_0^t \ha(s)\6s},
\end{equation}
where $C, \kappa$ are positive constants. 
\end{theorem}

This result shows that $x_t$ will leave a \nbh\ of size $\sigma
g(t)/\sqrt{2\ha(t)}$ of $\xhatdet_t$ typically after a time of order
$\eps/\ha(0)$. Once this \nbh\ has been left, the drift term starts
prevailing over the diffusion term, and one can show, (although this is not
trivial,) that the typical time needed to leave a \nbh\ of order one of the
saddle is of order $\eps\abs{\log\sigma}$. We will state a similar result
in Section~\ref{sec_bd} when discussing the dynamics after passing through
a pitchfork bifurcation point. 


\section{Stochastic resonance}
\label{sec_sr}

Up to now, we have considered situations in which the potential has bounded
curvature near its (isolated) extrema, so that for sufficiently small noise
intensities and not too long time scales, most paths are concentrated near
the bottom of the well they started in.

Not surprisingly, interesting phenomena occur when the condition on the
curvature is violated. Two cases can be considered:
\begin{itemiz}
\item	\defwd{Avoided bifurcation:} The potential well becomes flatter,
but the curvature does not vanish completely; sufficiently strong noise,
however, may drive solutions to another potential well. This mechanism is
responsible in particular for the phenomenon of \defwd{stochastic
resonance}.
\item	\defwd{Bifurcation:} The curvature at the bottom of the well
vanishes, say at time $0$; for $t>0$, new potential wells may be created
(e.\,g.\ pitchfork bifurcation) or not (e.\,g.\ saddle--node bifurcation).
\end{itemiz}

\noindent
We will discuss the possible phenomena in the case of a Ginzburg--Landau
potential
\begin{equation}
\label{sr1}
V(x,t) = \frac14 x^4 - \frac12 \mu(t) x^2 - \lambda(t)x.
\end{equation}
However, the precise form of the potential and the fact that it has at most
two wells are not essential. 

The potential \eqref{sr1} has two wells if $27\lambda^2 < 4\mu^3$ and one
well if $27\lambda^2 > 4\mu^3$. Crossing the lines $27\lambda^2 = 4\mu^3$,
$\mu>0$, corresponds to a saddle--node bifurcation, and crossing the
point $\lambda=\mu=0$ corresponds to a pitchfork bifurcation. Equilibrium
points are solutions of the equation $x^3-\mu(t)x-\lambda(t)=0$; we will
denote stable equilibria by $x^\star_\pm$, and the saddle, when present, by
$x^\star_0$. 

In this section, we investigate situations with avoided bifurcations, in
which the potential always has two wells, but the barrier between them
becomes low periodically. Bifurcation phenomena will be discussed in the
next two sections. 


\subsection{The mechanism of stochastic resonance}
\label{ssec_srm}

Let us consider the case where $\mu$ is a positive constant, say $\mu=1$,
and $\lambda(t)$ varies periodically, say $\lambda(t)=-A\cos(2\pi t)$. If
$\abs{\lambda}<\lc = 2/(3\sqrt3)$, then $V$ is a double-well potential. We
thus assume that $A<\lc$. 

In the absence of noise, the existence of the potential barrier prevents
the solutions from switching between potential wells. If noise is present,
but there is no periodic driving ($A=0$), solutions will cross the
potential barrier at random times, whose expectation is given by Kramers'
time $\eps\e^{2H/\sigma^2}$, where $H$ is the height of the barrier
($H=1/4$ in this case).

Interesting things happen when both noise and periodic driving $\lambda(t)$
are present. Then the potential barrier will still be crossed at random
times, but with a higher probability near the instants of minimal barrier
height (i.\,e., when $t$ is integer or half-integer). This phenomenon produces
peaks in the power spectrum of the signal, hence the name \defwd{stochastic
resonance} (SR). 

\begin{figure}
 \centerline{\psfig{figure=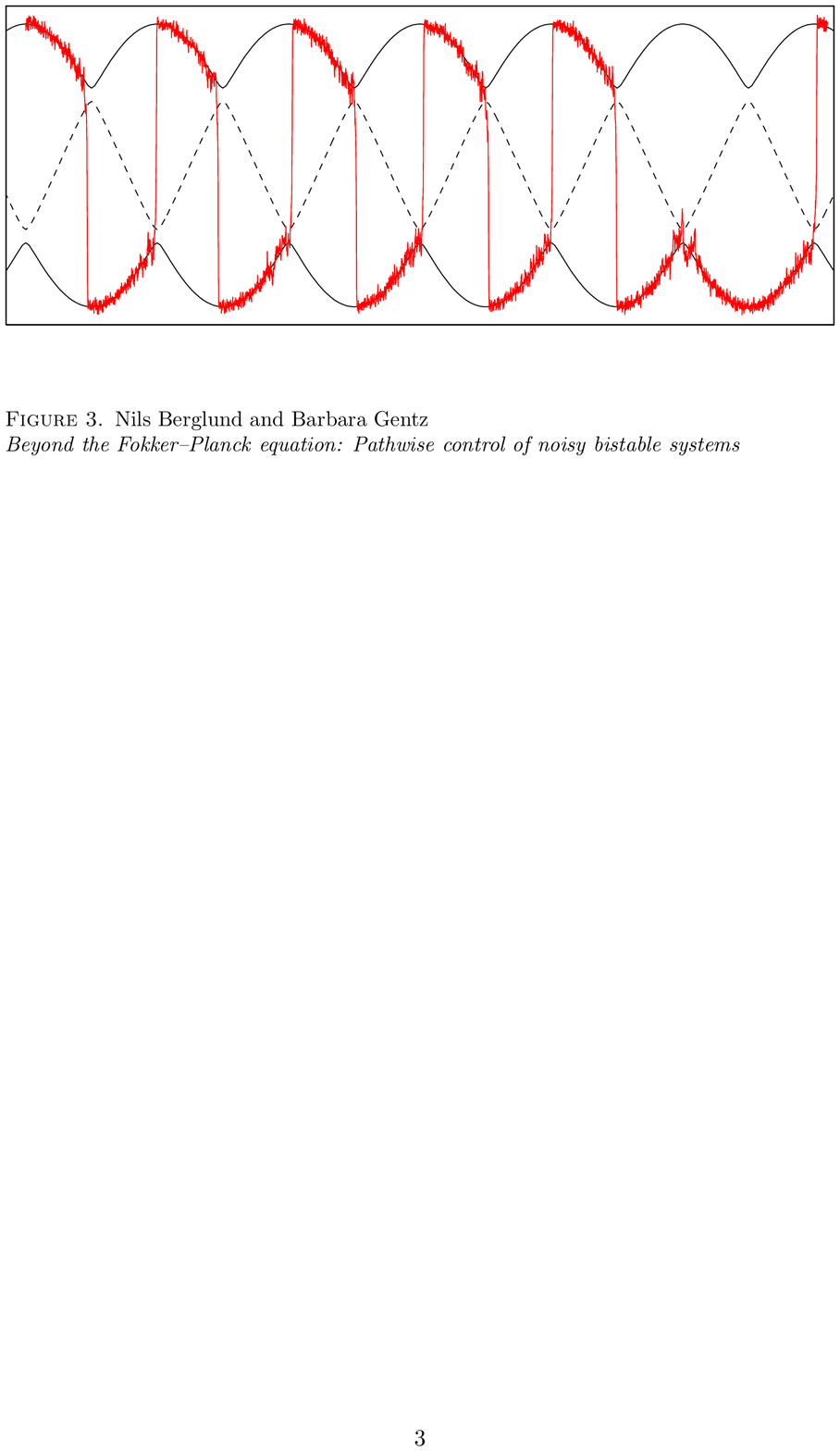,width=138mm,clip=t}}
 \captionspace
 \caption[]
 {A sample path of the equation of motion \eqref{srp1} in an asymmetrically
 perturbed double-well potential. Parameter values are $\eps=0.0025$,
 $a_0=0.005$ and $\sigma=0.065$, which is just above the threshold for
 transitions to be likely. The upper and lower full curves show the
 locations of the potential wells, while the broken curve marks the
 location of the saddle.}
\label{fig_sresasym}
\end{figure}

If the noise intensity is sufficiently large compared to the minimal
barrier height, transitions become likely twice per period (back and
forth), so that the signal $x_t$ is close, in some sense, to a periodic
function (\figref{fig_sresasym}). The amplitude of this oscillation may be
considerably larger than the amplitude of the forcing $\lambda(t)$, so that
the mechanism can be used to amplify weak periodic signals. This phenomenon
is also known as \defwd{noise-induced synchronization}
\cite{Neiman1,Neiman2}. Of course, too large noise intensities will spoil
the quality of the signal.  

The mechanism of stochastic resonance was originally introduced as a
possible explanation of the close-to-periodic appearance of the major
Ice Ages \cite{Benzi1,Benzi2}. Here the (quasi-)periodic forcing is
caused by variations in the Earth's orbital parameters (Milankovich factors),
and the additive noise models the fast unpredictable fluctuations
caused by the weather. Meanwhile, SR has been detected in a large
number of systems (see for instance~\cite{MW,WM,GHM} for reviews),
including ring lasers, electronic devices, and even the sensory system
of crayfish and paddlefish~\cite{Neiman-web}. 

Despite the many applications of SR, its mathematical description has
remained incomplete for two decades, although several limiting cases have
been studied in detail. The first approaches considered either potentials
that are piecewise constant in time \cite{Benzi1}, or two-level systems
with discrete space \cite{ET,McW}. Continuous time equations have been
mainly investigated through the Fokker--Planck equation, using methods from
spectral theory \cite{Fox,JH1} or linear response theory \cite{JH2}. The
main contribution of these approaches is an estimation of the
signal-to-noise ratio (SNR) of the power spectrum, as a function of the
noise intensity. The SNR is one of the possible quantitative measures of
the signal's periodicity, and behaves roughly like
$\e^{-H/\sigma^2}/\sigma^4$, which is maximal for $\sigma^2=H/2$. 
In~\cite{MS2}, an action functional is used to extend Kramers' result
to the case of small-amplitude forcing.

A description of individual paths has been given for the first time in
Freidlin's recent paper~\cite{Freidlin}. His results apply to a
general class of $n$-dimensional potentials, in the case where the
period $1/\eps$ of the driving scales like Kramers' time
$\e^{2H/\sigma^2}$. The fact that the minimal barrier height $H$ is
considered as constant while $\sigma$ tends to zero, implies that the
results only hold for exponentially long driving periods.  As
quantitative measure of the signal's periodicity, the
$L^p$-distance\footnote{The $L^p$-distance between $x_t$ and $\phi(t)$
is the integral of $\norm{x_t-\phi(t)}^p$ over a given time interval
(raised to the power $1/p$). Note that in contrast to a small supremum
norm, a small $L^p$-norm of $x_t-\phi(t)$ does not exclude that $x_t$
makes large excursions away from $\phi(t)$, as long as these
excursions are sufficiently short.}  between paths
$\set{x_t}_{t\geqs0}$ and a deterministic, periodic limit function
$\phi(t)$ is used. This limit function simply tracks the bottom of a
potential well, and jumps to the deeper well each time the potential
barrier becomes lowest. The $L^p$-distance is shown to converge to
zero in probability as $\sigma$ goes to zero. However, Freidlin's
techniques do not yield estimates on the speed of this convergence, or
its dependence on~$p$. 

Our techniques allow us to provide such estimates for one-dimensional
potentials, with a more natural distance than the $L^p$-distance: In fact,
we simply use a geometrical distance between paths and the limit function,
considered as curves in the $(t,x)$-plane. The analysis given below also
includes situations in which the minimal barrier height becomes small in
the small-noise limit. 


\subsection{Pathwise description}
\label{ssec_srp}

For the Ginzburg--Landau potential \eqref{sr1}, $\mu=1$ and
$\lambda(t)=-A\cos(2\pi t)$, the SDE takes the form 
\begin{equation}
\label{srp1}
\6x_t = \frac1\eps \bigbrak{x_t - x_t^3 - A\cos(2\pi t)}\6t +
\frac\sigma{\sqrt\eps}\6W_t.
\end{equation}
We assume that $A<\lc$, so that there are always two stable equilibria at
$x^\star_\pm(t)$ and a saddle at $x^\star_0(t)$. We introduce a parameter
$a_0=\lc-A$ which measures the minimal barrier height: At $t=0$, the
barrier height is of order $\smash{a_0^{3/2}}$ for small $a_0$, and the
distance between $x^\star_+$ and the saddle at $x^\star_0$ is of order
$\sqrt{a_0}$. At $t=\frac12$, the left-hand potential well at $x^\star_-$
is likewise close to the saddle. In order for transitions to become
possible on a time scale which is not exponentially large, we allow $a_0$
to become small with $\eps$.

Assume that we start at time $t_0=-1/4$ in the basin of attraction of the
right-hand potential well. Results from Section~\ref{sec_sw} show that
transitions are unlikely for $t\ll0$. Also, for $0\ll t\ll1/2$, paths will
be concentrated either near $x^\star_+$ or near $x^\star_-$. This allows us
to define the \defwd{transition probability} as 
\begin{equation}
\label{srp2}
\Ptrans = \bigprobin{t_0,x_0}{x_{t_1}<0}, 
\qquad\qquad
t_0 = -1/4, 
\quad
t_1 = 1/4.
\end{equation}
The properties of $\Ptrans$ do not depend sensitively of the choices of
$t_0$ and $t_1$, as long as $-1/2\ll t_0\ll 0\ll t_1\ll 1/2$. Also the
level $0$ can be replaced by any level lying between $x^\star_-(t)$ and
$x^\star_+(t)$ for all $t$. Denoting by $a\vee b$ the maximum of two real
numbers $a$ and $b$, our main result can be formulated as follows. 

\begin{figure}
 \centerline{\psfig{figure=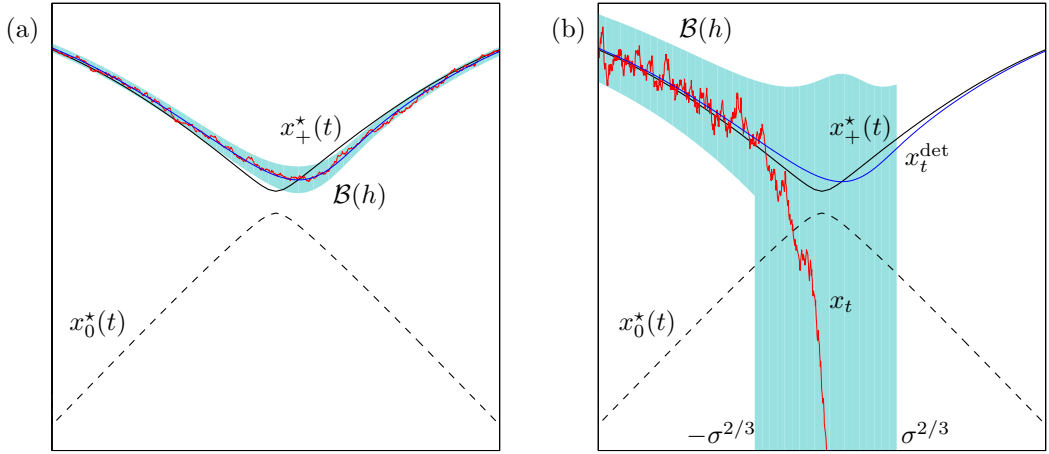,height=60mm,clip=t}}
 \captionspace
 \caption[]
 {Sample paths of Equation~\eqref{srp1} in a \nbh\ of time $0$, when the
 barrier height is minimal, for two different noise intensities. Full
 curves mark the location $x^\star_+(t)$ of the right-hand potential well,
 broken curves the location $x^\star_0(t)$ of the saddle. Parameter values
 are $\eps=0.0125$, $a_0=0.002$ and (a) $\sigma=0.012$, (b) $\sigma=0.07$.
 In case (a), the path remains in the set $\cB(h)$, shown here for $h=3$,
 which is centred at the deterministic solution $\xdet_t$. In case (b),
 $x_t$ remains in $\cB(h)$ as long as the width of $\cB(h)$ is smaller than
 the distance between $\xdet_t$ and the saddle, that is, for
 $t\ll-\sigma^{2/3}$. The path jumps to the left-hand potential well
 during the time interval $[-\sigma^{2/3},\sigma^{2/3}]$.}
\label{fig_sresasymdetail}
\end{figure}

\begin{theorem}[{{\rm\cite[Theorems~2.6 and~2.7]{BG2}}}]
\label{thm_srp}
For the noise intensity, there is a threshold level 
$\sigmac = (a_0\vee\eps)^{3/4}$ with the following properties:
\begin{enum}
\item	If $\sigma<\sigmac$, then 
\begin{equation}
\label{srp3}
\Ptrans \leqs \frac C\eps \e^{-\kappa\sigmac^2/\sigma^2}
\end{equation}
for some $C,\kappa>0$. Paths are concentrated in a strip of width 
$\sigma/(\sqrt{\abs t}\vee \sigmac^{1/3})$ around the deterministic
solution tracking $x^\star_+(t)$ (\figref{fig_sresasymdetail}a). 

\item	If $\sigma>\sigmac$, then 
\begin{equation}
\label{srp4}
\Ptrans \geqs 1 - C \e^{-\kappa \sigma^{4/3}/(\eps\abs{\log\sigma})}
\end{equation}
for some $C,\kappa>0$. Transitions are concentrated in the interval
$[-\sigma^{2/3},\sigma^{2/3}]$. Moreover, for $t\leqs-\sigma^{2/3}$, paths
are concentrated in a strip of width  $\sigma/\sqrt{\abs t}$ around the
deterministic solution tracking $x^\star_+(t)$, while for
$t\geqs\sigma^{2/3}$, they are concentrated in a strip of width 
$\sigma/\sqrt{t}$ around a deterministic solution tracking $x^\star_-(t)$
(\figref{fig_sresasymdetail}b).
\end{enum}
\end{theorem}

The crossover is quite sharp: For $\sigma\ll\sigmac$, transitions between
potential wells are very unlikely, while for $\sigma\gg\sigmac$, they are
very likely. By \lq\lq concentrated in a strip of width $w$\rq\rq, we mean
that the probability that a path leaves a strip of width $hw$ decreases
like $\e^{-\kappa h^2}$ for some $\kappa>0$. The \lq\lq typical width
$w$\rq\rq\ is our measure of the deviation from the deterministic periodic
function, which tracks one potential well in the small-noise case, and
switches back and forth between the wells in the large-noise case. 

Theorem~\ref{thm_srp} implies in particular that for the periodic signal's
amplification to be optimal, the noise intensity $\sigma$ should exceed the
threshold $\sigmac$. Larger noise intensities will increase both spreading
of paths (especially just before they cross the potential barrier) and size
of transition window, and thus spoil the output's periodicity. 

It has been proposed to identify stochastic resonance and
synchronization with a phase-locking mechanism \cite{Neiman2}, but the
definition of a phase remained problematic. The fact that the majority
of paths are contained in small space--time sets makes it possible to
associate a random phase with those paths. For instance, when
$\sigma>\sigmac$, most paths circle the origin of the
$(\lambda,x)$-plane counterclockwise (see
\figref{fig_hystcycles}c). For all paths not containing the origin,
one can define a random phase $\ph_t$ by
$\tan(\ph_t)=x_t/\lambda(t)$. Although $\ph_t$ fluctuates, it is
likely to increase at an average rate of $2\pi$ per cycle. 

Part of our results should appear quite natural. If $a_0>\eps$, the
threshold noise level $\sigmac=\smash{a_0^{3/4}}$ behaves like the square
root of the minimal barrier height, which is consistent with the SNR being
optimal for $\sigma^2=H/2$. However, $\sigmac$ saturates at $\eps^{3/4}$
for all $a_0\leqs\eps$. Hence, even driving amplitudes arbitrarily close to
$\lc$ cannot increase the transition probability. This is a rather subtle
dynamical effect, mainly due to the fact that even if the barrier vanishes
at $t=0$, it is lower than $\eps^{3/2}$ during too short a time interval
for paths to take advantage. The situation is the same as if there were an
\lq\lq effective potential barrier\rq\rq\ of height proportional to
$\sigmac^2$. 

Another remarkable fact is that for $\sigma>\sigmac$, neither the
transition probability nor the width of the transition windows depend on
the driving amplitude to leading order. In the remainder of this
subsection, we are going to explain some ideas of the proof of
Theorem~\ref{thm_srp}, which will hopefully clarify some of the above
surprising properties.  

The first step is to understand the behaviour of the solutions of
\eqref{srp1} in the deterministic case $\sigma=0$. This problem belongs to
the field of dynamical bifurcations, the theory of which is relatively well
developed. We follow here a framework allowing to determine scaling laws of
solutions near bifurcation points, which has been presented in \cite{BK}.
The main idea is that when approaching a bifurcation point
$(t^\star,x^\star)$, the distance between the adiabatic solution and the
static equilibrium branch scales like $\eps/\abs{t-t^\star}^\rho$ for
$t\leqs t^\star-\eps^\nu$, and like $\eps^\mu$ for
$\abs{t-t^\star}\leqs\eps^\nu$. The rational numbers $\rho$, $\nu$ and
$\mu=1-\rho\nu$ are universal exponents, which can be deduced from the
Newton polygon of the bifurcation point. 

Tihonov's theorem implies that away from the avoided bifurcation point at
$t=0$, solutions $\xdet_t$ of the deterministic equation track the
equilibrium branch $x^\star_+(t)$ adiabatically at a distance of order
$\eps$. It is thus sufficient to understand what happens in a small \nbh\
of the almost-bifurcation. Note that if $\lambda=\lc$, the right-hand well
and the saddle merge at $x=1/\sqrt3$. It is thus helpful to consider the
translated variable $\ydet_t = \xdet_t-1/\sqrt3$, which obeys the
differential equation
\begin{equation}
\label{srp5}
\eps\dot y = ct^2 - \sqrt3 y^2 + a_0 + \text{higher order terms}, 
\qquad\qquad
c = 2\pi^2\lc. 
\end{equation}
Consider the \lq\lq worst\rq\rq\ case $a_0=0$. Then the right-hand side of
\eqref{srp5} describes a transcritical bifurcation between the equilibria
$y^\star_{+,0} = \pm\sqrt c 3^{-1/4}\abs{t} + \Order{t^2}$. One shows that
the solution $\ydet_t$ tracks $y^\star_+(t)$ at a distance scaling like
$\eps/\abs{t}$ for $\abs{t}\geqs\sqrt\eps$, and like $\sqrt\eps$ for
$\abs{t}\leqs\sqrt\eps$. In fact, $\ydet_t$ never approaches the saddle
closer than a distance of order $\sqrt\eps$, which is because the term
$-\sqrt3 y^2$ only dominates during a short time interval of order
$\sqrt\eps$. 

The same qualitative behaviour holds for $0<a_0\leqs\eps$. For $a_0>\eps$,
one can show that $\xdet_t$ tracks $x^\star_+(t)$ at a distance never
exceeding $\Order{\eps/\sqrt{a_0}}$. Since $x^\star_+(0)-x^\star_-(0)$ is of
order $\sqrt{a_0}$, $\xdet_t$ never approaches the saddle closer than a
distance of order $\sqrt{a_0}$. 

Let us now consider the random motion near $\xdet$ for $\sigma>0$. We
denote, as usual, by $a(t)$ the curvature of the potential at the
deterministic solution $\xdet_t$. By \eqref{srp5}, $a(t)$ behaves, near
$t=0$, like $-\ydet_t$, which we know to behave like
$-(\abs{t}\vee\sqrt{a_0})$ for $a_0>\eps$, and like
$-(\abs{t}\vee\sqrt\eps)$ for $a_0\leqs\eps$. It turns out that
Theorem~\ref{thm_swn} can be extended to the present situation, to show
that paths are concentrated in a strip around $\xdet_t$. The width of this
strip is again related to the standard deviation of a linearized process,
and behaves like 
\begin{equation}
\label{srp6}
\frac\sigma{\sqrt{\abs{a(t)}}} \asymp 
\frac\sigma{\sqrt{\abs t}\vee \sigmac^{1/3}}. 
\end{equation} 
However, this property only holds under the condition that the spreading is
always smaller than the distance between $\xdet_t$ and the saddle, which
scales like $\abs{t}\vee\smash{\sigmac^{2/3}}$. We thus have to require
$\sigma<\abs{t}^{3/2}\vee\sigmac$, so that 
\begin{itemiz}
\item	if $\sigma<\sigmac$, the condition is always satisfied, and thus
\eqref{srp3} follows from the generalization of Theorem~\ref{thm_swn} with
$h=\sigmac/\sigma$ (\figref{fig_sresasymdetail}a);
\item	if $\sigma>\sigmac$, the condition is only satisfied for
$t\leqs-\sigma^{2/3}$ (\figref{fig_sresasymdetail}b). 
\end{itemiz}

It thus remains to understand what happens for $t\geqs-\sigma^{2/3}$ if
$\sigma>\sigmac$. Here the main idea is that during the time interval
$[-\sigma^{2/3},\sigma^{2/3}]$, the process $x_t$ has a certain number of
trials to reach the saddle. If $x_t$ reaches the saddle, it has roughly
probability $1/2$ to move in each direction. If it moves far enough towards
the left, it will most probably fall into the left-hand potential well, and
is unlikely to come back for the remaining half-period. If it moves to the
right, it has failed to overcome the barrier, but can try again during the
next excursion. 

One can define a typical time $\Delta t$ for an excursion as the time needed
for a path starting near $x^\star_+$ to reach and overcome the barrier with
non-negligible probability, say with probability $1/3$. One can show, by
comparison with suitable linearized processes, that this typical time is
determined by the condition 
\begin{equation}
\label{srp7}
\abs{\alpha(t,t+\Delta t)} = \int_t^{t+\Delta t} \abs{a(s)}\6s 
= \text{{\it const }} \eps\abs{\log\sigma}.
\end{equation}
To obtain this, one first checks that the curvature of the potential is the
same, up to sign reversal, at the adiabatic solutions tracking the bottom
of the well and the saddle. To overcome the saddle, the integral in
\eqref{srp7} must be of order $\eps$, similarly as in Theorem~\ref{thm_es}.
The factor $\abs{\log\sigma}$ is needed to reach a distance of order $1$
from the saddle. 
Now the maximal number $N$ of trials during the interval
$[-\sigma^{2/3},\sigma^{2/3}]$ is given by 
\begin{equation}
\label{srp8}
N = {\text{{\it const }}}
\frac{\abs{\alpha(\sigma^{2/3},-\sigma^{2/3})}}{\eps\abs{\log\sigma}}
\geqs \text{{\it const }} \frac{\sigma^{4/3}}{\eps\abs{\log\sigma}}.
\end{equation}
Finally, the Markov property implies that the probability {\it not\/} to
overcome the saddle during $N$ trials is bounded by 
\begin{equation}
\label{srp9}
\Bigpar{\frac23}^N = \exp\biggset{-N\log\Bigpar{\frac23}} 
\leqs \exp\biggset{-\text{{\it const }}
\frac{\sigma^{4/3}}{\eps\abs{\log\sigma}}},
\end{equation}
which proves \eqref{srp4}. An important point to note is that the transition
probability is not determined by the curvature of the potential at the
saddle, but by the curvature at the deterministic solution tracking the
saddle, which may be larger for small $a_0$. 


\subsection{Symmetric potentials}
\label{ssec_srs}

Another case of interest is the Ginzburg--Landau potential \eqref{sr1} with
$\lambda\equiv0$ and $\mu(t) = a_0 + 1 - \cos(2\pi t)$, $a_0>0$. Then
$V(x,t)$ is always symmetric, with minima at
$x^\star_\pm(t)=\pm\sqrt{\mu(t)}$ and a barrier height $\frac14\mu(t)^2$
becoming small at integer times. The associated SDE is 
\begin{equation}
\label{srs1}
\6x_t = \frac1\eps \bigbrak{(a_0 + 1 - \cos(2\pi t))x_t - x_t^3} \6t +
\frac{\sigma}{\sqrt\eps} \6W_t.
\end{equation}
As before, transitions between potential wells are most likely when the
barrier is lowest. We can thus define a transition probability as in
\eqref{srp2}, with $-1\ll t_0\ll 0\ll t_1\ll 1$. We again assume that the
process starts in the right-hand well.  In this case, the result
corresponding to Theorem~\ref{thm_srp} is 

\begin{figure}
 \centerline{\psfig{figure=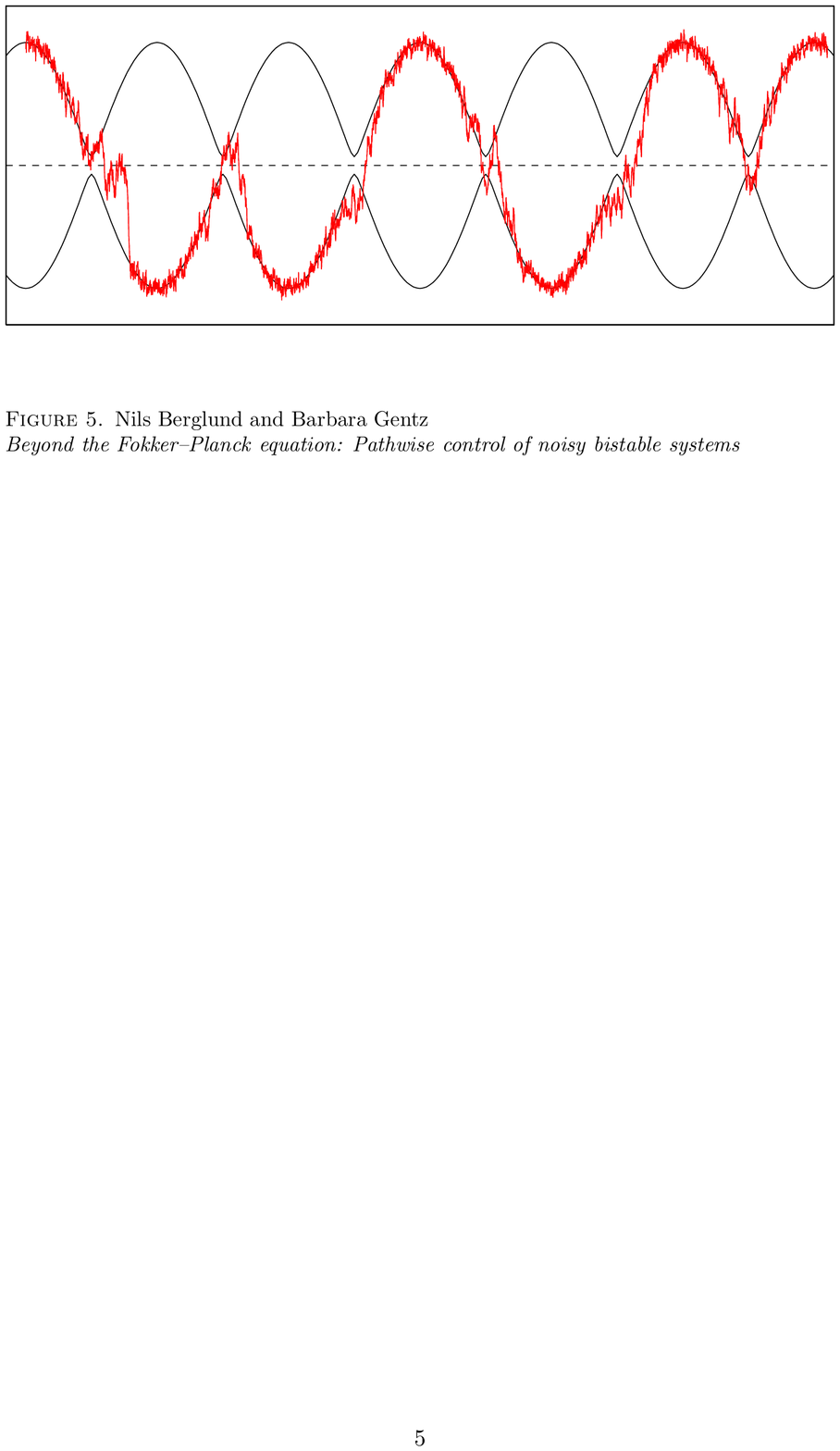,width=138mm,clip=t}}
 \captionspace
 \caption[]
 {A sample path of the equation of motion \eqref{srs1} in a periodically
 modulated symmetric double-well potential. Parameter values are
 $\eps=0.005$, $a_0=0.005$ and $\sigma=0.075$, which is above the threshold
 for transitions to occur with probability close to $1/2$. The upper and
 lower full curves show the locations of the potential wells, while the
 broken line marks the location of the saddle.}
\label{fig_sressym}
\end{figure}

\begin{theorem}[{{\rm\cite[Theorems~2.2--2.4]{BG2}}}]
\label{thm_srs}
There is a threshold noise level $\sigmac = a_0\vee\eps^{2/3}$ with the
following properties:
\begin{enum}
\item	If $\sigma<\sigmac$, then 
\begin{equation}
\label{srs2}
\Ptrans \leqs \frac C\eps \e^{-\kappa\sigmac^2/\sigma^2}
\end{equation}
for some $C,\kappa>0$. Paths are concentrated in a strip of width 
$\sigma/(\abs t\vee \sqrt{\sigmac}\mskip1.5mu)$ around the
deterministic solution tracking $x^\star_+(t)$. 

\item	If $\sigma>\sigmac$, then 
\begin{equation}
\label{srs3}
\Ptrans \geqs \frac12 - C \e^{-\kappa \sigma^{3/2}/(\eps\abs{\log\sigma})}
\end{equation}
for some $C,\kappa>0$. Transitions are concentrated in the interval
$[-\sqrt{\sigma},\sqrt{\sigma}\mskip2mu]$. Moreover, for
$t\leqs-\sqrt{\sigma}$, paths are concentrated in a strip of width 
$\sigma/\abs t$ around the deterministic solution tracking $x^\star_+(t)$,
while for $t\geqs\sqrt{\sigma}$, they are concentrated in a strip of width 
$\sigma/t$ around a deterministic solution tracking either $x^\star_+(t)$
or $x^\star_-(t)$.
\end{enum}
\end{theorem}

The main difference with respect to the previous case is that due to the
symmetry, $\Ptrans$ can never exceed $1/2$. The limiting process obtained
by letting $\sigma$ go to zero but keeping $\sigma>\sigmac$ is no longer a
deterministic function, but a \lq\lq Bernoulli\rq\rq\ process, choosing
between the left and the right potential well with probability $1/2$ at
integer times (\figref{fig_sressym}). 

Another difference lies in the distribution of barrier crossing times in
the transition window. In the asymmetric case, paths may overcome the
saddle as soon as $t\geqs-\sigma^{2/3}$, and are unlikely to return to the
shallower well. In the symmetric case, paths may jump back and forth
between both wells up to time $\sqrt\sigma$, before settling for a
potential well. 


\subsection{Modulated noise intensity}
\label{ssec_sri}

Other mechanisms leading to stochastic resonance have been examined, for
instance periodic forcing which is not deterministic, but affects the
noise intensity, a situation arising in power amplifiers \cite{Dykman}.
This case can be analyzed by the same method as the previous ones, but the
results are quite different. 

Let us consider the motion in a static symmetric double-well potential,
described by the SDE
\begin{equation}
\label{sri1}
\6x_t = \frac1\eps \bigbrak{\mu_0 x_t - x_t^3}\6t + \frac\sigma{\sqrt\eps}
g(t) \6W_t,
\end{equation}
where $\mu_0>0$ is fixed and $g(t)$ is periodic. We assume that $g(t)\geqs
\eps\abs{\log\sigma}/\mu_0$ for all $t$. Theorem~\ref{thm_swn} shows
that for sufficiently weak noise, paths starting at time $t_0$ at the bottom
$\sqrt{\mu_0}$ of the right-hand potential well remain concentrated in
a strip around $\sqrt{\mu_0}$, with width proportional to $\sigma
g(t)/(2\sqrt{\mu_0})$. This holds as long as the spreading is smaller than a
constant times the distance $\sqrt{\mu_0}$ between well and saddle. 
The probability to cross the saddle before time $t$ is bounded by  
\begin{equation}
\label{sri2}
P(t) = C(t,\eps) \exp\biggset{-\frac\kappa{\sigma^2}
\biggpar{\frac{2\mu_0}{\hg(t)}}^2}, 
\qquad
\text{where}
\qquad
\hg(t) = \sup_{t_0\leqs s\leqs t} g(s)
\end{equation}
and $\kappa>0$. Thus if $g$ reaches its maximum $\hg(0)$ at time $0$,
the probability to see a transition during one period satisfies 
\begin{equation}
\label{sri3}
\Ptrans \leqs C(1,\eps) \e^{-\kappa \sigmac^2/\sigma^2}
\qquad
\text{for}
\qquad
\sigma\leqs\sigmac = \frac{2\mu_0}{\hg(0)}. 
\end{equation}
Note that here, as the potential is static, the threshold value for the
noise intensity can be guessed from Kramers' time, assuming constant
$g$. Taking into account that $g$ is {\it not\/} necessarily
constant, we see that for $\sigma>\sigmac$, transitions are likely to
happen in the time interval during which $\sigma g(t) \geqs 2\mu_0$. For
instance, if $g(t)$ behaves quadratically near its unique maximum,
this transition window is given by  
\begin{equation}
\label{sri4}
t^2 \leqs \text{{\it const }} g(0) \Bigpar{1-\frac\sigmac\sigma}. 
\end{equation}
In contrast to the previous cases, however, the transition times are less
concentrated, in the sense that for times $t_0<t_1<t_2$ before the 
transition window, 
\begin{equation}
\label{sri5}
P(t_1) \simeq P(t_2)^{(\hg(t_2)/\hg(t_1))^2}.
\end{equation}
A similar argument as in the previous cases shows that for $\sigma>\sigmac$,
\begin{equation}
\label{sri6}
\Ptrans \geqs \frac12 - \text{{\it const }}  \exp\biggset{-\kappa
\frac{2\mu_0\Delta}{\eps\abs{\log\sigma}}},
\end{equation}
where $\Delta$ is the length of the transition window. 

If the potential is made asymmetric, so that a constant term $\lambda_0>0$
is added to the drift term in \eqref{sri1}, the critical noise intensities
needed to reach the saddle from the shallower left-hand well and the deeper
right-hand well will be different (one can check that their ratio is
$1+\Order{\lambda_0/\smash{\mu_0^{3/2}}}$). As a consequence, transitions
from the shallower to the deeper well will be likely as soon as the noise
intensity exceeds the smaller threshold, while transitions in both
directions are likely when the noise intensity exceeds the larger
threshold. When the noise intensity drops below the smaller threshold again,
$x_t$ will be in the deeper right-hand well with larger probability. The net
effect is that $x_t$ will visit both wells near integer times, but has 
only small probability to remain in the shallow well after the
transition window. Thus the periodic signal is not amplified in the
same way as discussed before, but nevertheless we observe an
amplification mechanism which allows to read off at which times the
threshold is exceeded. 


\section{Hysteresis}
\label{sec_hy}

Hysteresis is another characteristic phenomenon of bistable systems. Let us
consider again the motion in the Ginzburg--Landau potential \eqref{sr1}
with $\mu\equiv1$ and $\lambda(t)=-A\cos(2\pi t)$, but without imposing the
restriction $A<\lc$. In the deterministic case $\sigma=0$ the equation of
motion reads
\begin{equation}
\label{hy1}
\eps\dtot{x_t}t = x_t - x_t^3 + \lambda(t).
\end{equation}
We may ask the question: How does $x_t$ behave, as a function of
$\lambda(t)$, in the adiabatic limit $\eps\to0$? Intuitively, $x_t$ will
always track the bottom of a potential well. A naive way to see this is to
set formally $\eps$ equal to zero in \eqref{hy1}: We obtain the algebraic
equation $x-x^3+\lambda=0$, which admits three branches of solutions (see
\figref{fig_hystdet}); $X^\star_+(\lambda)$ and $X^\star_-(\lambda)$
correspond to potential wells, and $X^\star_0(\lambda)$ to a saddle, which
exists only for $\abs{\lambda}<\lc$. 

If the amplitude $A$ is smaller than the critical value $\lc$, there are
always two potential wells separated by a barrier. Hence $x_t$ will always
track the bottom of the same well in the limit $\eps\to0$, so that the
instantaneous value of $\lambda$ is sufficient to determine the state
(provided we know in which potential well the process started). 

If $A$ is larger than $\lc$, however, a saddle--node bifurcation point is
crossed whenever $\abs{\lambda}$ reaches $\lc$ from below: The potential
well tracked by $x_t$ disappears, so that $x_t$ jumps to the other well,
which is unaffected by the bifurcation (\figref{fig_potential}). As a
result, the state $x_t$ is not uniquely defined by the instantaneous value
of $\lambda$ if $\abs{\lambda}\leqs\lc$: $x_t$ tracks the bottom
$X^\star_-(\lambda(t))$ of the left-hand well if $\lambda$ increases, and
the bottom $X^\star_+(\lambda(t))$ of the right-hand well if $\lambda$
decreases. This phenomenon is called \defwd{hysteresis}. The hysteresis
cycle consists of the branches $X^\star_\pm(\lambda)$, $\abs\lambda \leqs
A$, and two vertical lines on which $\abs\lambda=\lc$. It encloses an area
$\cA_0=3/2$, called \defwd{static hysteresis area.} In many applications,
$x$ and $\lambda$ are thermodynamically conjugated variables, and the
hysteresis area represents the energy dissipation per cycle.  


\subsection{Dynamical hysteresis and scaling laws}
\label{ssec_hds}

\begin{figure}
 \centerline{\psfig{figure=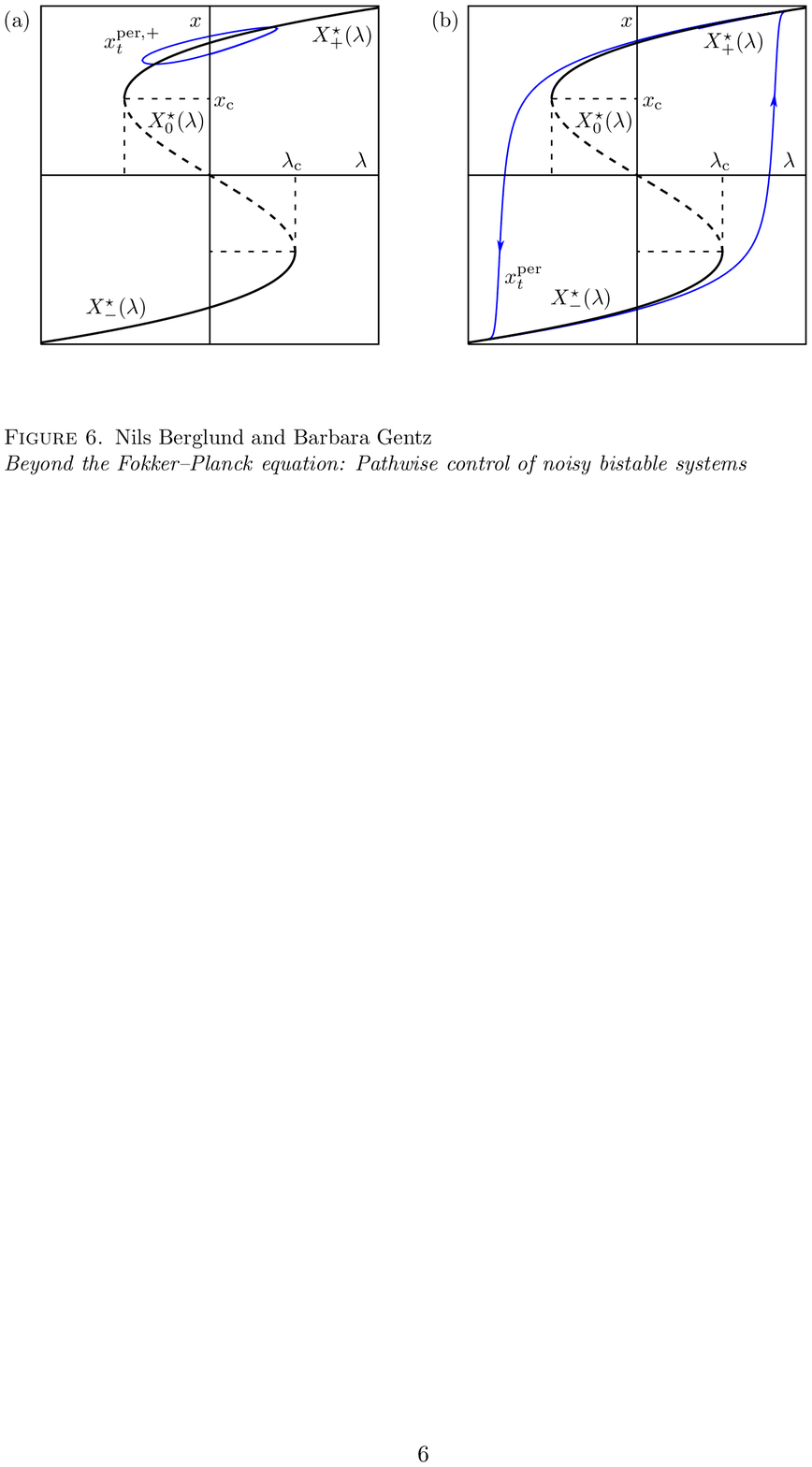,height=50mm,clip=t}}
 \captionspace
 \caption[]
 {Periodic solutions of the deterministic equation~\eqref{hy1}, (a) in a
 case where the amplitude $A$ of $\lambda(t)$ is smaller than $\lc$, and
 (b) in a case where it is larger than $\lc$. The enclosed area scales like
 $\eps A$ in case (a), and like $\cA_0+\eps^{2/3}(A-\lc)^{1/3}$ in case
 (b), where $\cA_0$ is the static hysteresis area. Potential wells
 $X^*_\pm(\lambda)$ are displayed as full curves, the saddle
 $X^\star_0(\lambda)$ as a broken curve.}
\label{fig_hystdet}
\end{figure}

Consider now what happens when $\eps$ is small but positive. The solutions
of Equation~\eqref{hy1} will not react instantaneously to changes in the
potential, so that the shape of hysteresis cycles is modified. It is
important to understand the $\eps$-dependence of quantities such as the
average of $x_t$ over one period, the value of $\lambda$ when $x_t$ changes
sign, and the area enclosed by the hysteresis cycle. It is known that there
are constants $\gamma_1>\gamma_0>0$ such that the following properties hold:
\begin{itemiz}
\item	If $A\leqs\lc+\gamma_0\eps$, then $x_t$ cannot change sign (except
during the very first period, if the process does not start near the bottom
of a well). There exist two stable periodic orbits, one tracking each
potential well (\figref{fig_hystdet}a). Each encloses an area $\cA$
satisfying
\begin{equation}
\label{hds1}
\cA \asymp A\eps,
\end{equation} 
and the average of $x_t$ over each cycle is nonzero. The notation $\asymp$
is a shorthand to indicate that $c_-A\eps\leqs\cA\leqs c_+A\eps$ for some
constants $c_\pm>0$ independent of $\eps$ and $A$. 

\item	If $A\geqs\lc+\gamma_1\eps$, then $x_t$ changes sign twice per
period. All orbits are attracted by the same periodic orbit
(\figref{fig_hystdet}b), corresponding to a hysteresis cycle with zero
average and area $\cA$ satisfying
\begin{equation}
\label{hds2}
\cA - \cA_0 \asymp \eps^{2/3} (A-\lc)^{1/3}.
\end{equation}
When $x_t$ changes sign, the parameter $\lambda$ satisfies
$\abs{\lambda}-\lc \asymp \eps^{2/3} (A-\lc)^{1/3}$.  

\item	If $\lc+\gamma_0\eps < A <\lc+\gamma_1\eps$, several hysteresis
cycles may coexist, some of them satisfying \eqref{hds1} and others
satisfying \eqref{hds2}. 
\end{itemiz}

\noindent
The scaling law \eqref{hds2} was first derived in \cite{Jung} for $A-\lc$ of
order $1$, where Equation~\eqref{hy1} was used to model a bistable laser.
The case where $A$ is close to $\lc$ has been analysed in \cite{BK}. 

An equation qualitatively similar to \eqref{hy1} describes the dynamics of a
Curie$\mskip1.5mu$--$\mskip-1.5mu$Weiss model of a ferromagnet,
subject to a periodic magnetic field $\lambda(t)$, in the limit of
infinitely many spins \cite{Martin}. The transition between the small
and large amplitude regimes has been called \lq\lq dynamic phase
transition\rq\rq\ in \cite{TO}.  

The magnetization obeys a deterministic differential equation only in the
limit of infinite system size. The effect of the number $N$ of spins being
finite can be modeled, in first approximation, by an additive white noise of
intensity proportional to $1/\sqrt N$ \cite{Martin}. It is thus of major
importance to understand the effect of additive noise on the properties of
hysteresis cycles. 

Langevin equations have already been studied for multi-dimensional
Ginzburg--Landau potentials. Then, however, the mechanism leading to
hysteresis is different, because there is no potential barrier between
stable states. Numerical simulations \cite{RKP} suggested that the
area of hysteresis cycles should follow the scaling law
$\cA\asymp\eps^{1/3}A^{2/3}$, while various theoretical arguments
indicate that $\cA\asymp\eps^{1/2}A^{1/2}$ \cite{DT,SD,ZZ}. It is not
clear whether such a scaling law exists for the Ising model
\cite{SRN}. 

For clarity, we will keep interpreting $x_t$ as magnetization and
$\lambda(t)$ as magnetic field. There exist, however, many other instances
where the dynamics is described by a periodically forced Langevin equation.
For instance, in models for the Atlantic thermohaline circulation, $x_t$
represents the salinity difference between high and middle latitude, and
$\lambda(t)$ represents the atmospheric freshwater flux \cite{Stommel,Rah}.
The effect of additive noise on this system has been investigated, for
instance in \cite{Cessi}, while the properties of hysteresis cycles
were considered in particular in \cite{Monahan}. 

The fact that additive noise may create relaxation oscillations has
been discussed in~\cite{Freidlin2}, where the motion of a light
particle in a randomly perturbed field is investigated with the help
of large deviation theory.


\subsection{The effect of additive noise}
\label{ssec_han}

We consider the Langevin equation 
\begin{equation}
\label{han1}
\6x_t = \frac1\eps \bigbrak{x_t - x_t^3 - A\cos(2\pi t)}\6t +
\frac\sigma{\sqrt\eps}\6W_t,
\end{equation}
where $A>0$. We denote $A-\lc$ by $a_0$, but in contrast to
Section~\ref{sec_sr} (where $a_0$ had opposite sign), we do not impose that
$a_0$ is a small parameter, and we allow positive as well as negative
$a_0$. Let us fix a deterministic initial condition $(t_0=-1/2,x_0>0)$,
such that the solution $\xdet_t$ of the deterministic equation \eqref{hy1}
with $\xdet_{t_0}=x_0$ is attracted by the right-hand potential well. 

We are interested in the quantity
\begin{equation}
\label{han2}
\cA(\eps,\sigma) = -\int_{-1/2}^{1/2} x_t \lambda'(t)\6t
\end{equation}
measuring the area enclosed by $x_t$ in the $(\lambda,x)$-plane during one
period ($x_t$ does not necessarily form a closed loop, but $\cA$ still
represents the energy dissipation). $\cA(\eps,0)$ is the area enclosed by
$\xdet_t$, and behaves like \eqref{hds1} or \eqref{hds2}. For positive
$\sigma$, $\cA(\eps,\sigma)$ is a random variable, the distribution of
which we want to characterize. 

Another random quantity of interest is the value $\lambda^0$ of the magnetic
field when $x_t$ changes sign for the first time:
\begin{equation}
\label{han3}
\lambda^0 = \lambda(\tau^0), \qquad\qquad
\tau^0 = \inf\bigsetsuch{t>t_0}{x_t\leqs0}.
\end{equation}

Results from Section~\ref{sec_sr} already allow us to make some predictions
for the case $a_0<0$. For $\sigma\ll\sigmac = (\abs{a_0}\vee\eps)^{3/4}$,
$x_t$ is unlikely to switch between potential wells, so that
$\cA(\eps,\sigma)$ will be concentrated near the deterministic value
$\cA(\eps,0)$, which is of order $\eps$ (\figref{fig_hystcycles}a). For
$\sigma\gg\sigmac$, $x_t$ is likely to cross the potential barrier at a
random time $\tau^0$ which behaves typically like $-\sigma^{2/3}$. The
corresponding field $\lambda^0$ behaves like $\lc-\sigma^{4/3}$
(\figref{fig_hystcycles}c). Thus additive noise of sufficient intensity
will lead to a hysteresis area which is smaller, by an amount of order
$\sigma^{4/3}$, than the static area $\cA_0$. 

The same behaviour can be shown to hold for positive $a_0$ up to order
$\eps$. In this case, the potential barrier vanishes during a short time
interval, which is too short, however, for $x_t$ to notice.  For
$a_0>\eps$, there is a similar transition between a small-noise regime
(\figref{fig_hystcycles}b), where $x_t$ is likely to track the
deterministic solution, and a large-noise regime, where it typically
crosses the potential barrier some time before the barrier vanishes. The
threshold value of $\sigma$ delimiting both regimes is again deduced from
the variance of the equation linearized around $\xdet$, and  turns out to
be $\sigmac = \esa0^{1/2}$. For $\sigma>\sigmac$, the typical value
$\lambda^0$ of the field when $x_t$ changes sign is again found to behave
like $\lc-\sigma^{4/3}$.

\begin{figure}
 \centerline{\psfig{figure=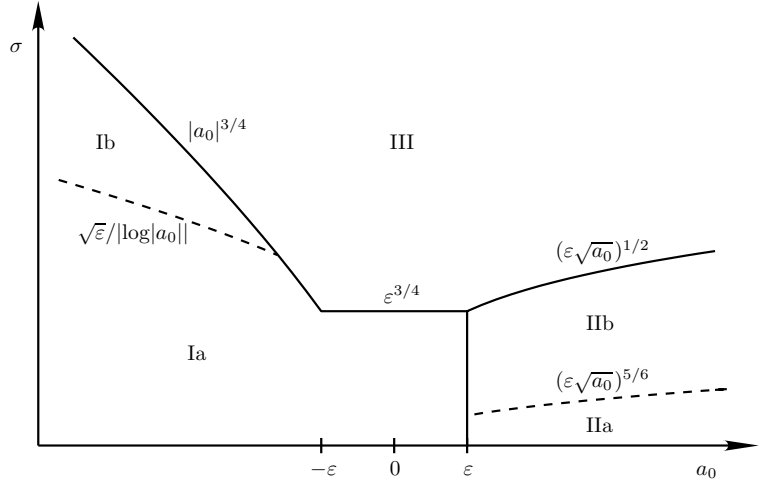,width=100mm,clip=t}}
 \captionspace
 \caption[]
 {Definition of the parameter regimes for hysteresis cycles, shown in the
 plane $(a_0=A-\lc,\sigma)$ for a fixed value of $\eps$. The behaviour of
 the hysteresis area $\cA(\eps,\sigma)$ in each regime is described in
 Theorem~\ref{thm_han}. Typical hysteresis cycles are shown in
 \figref{fig_hystcycles}.}
\label{fig_hystdiagram}
\end{figure}

\begin{figure}
 \centerline{\psfig{figure=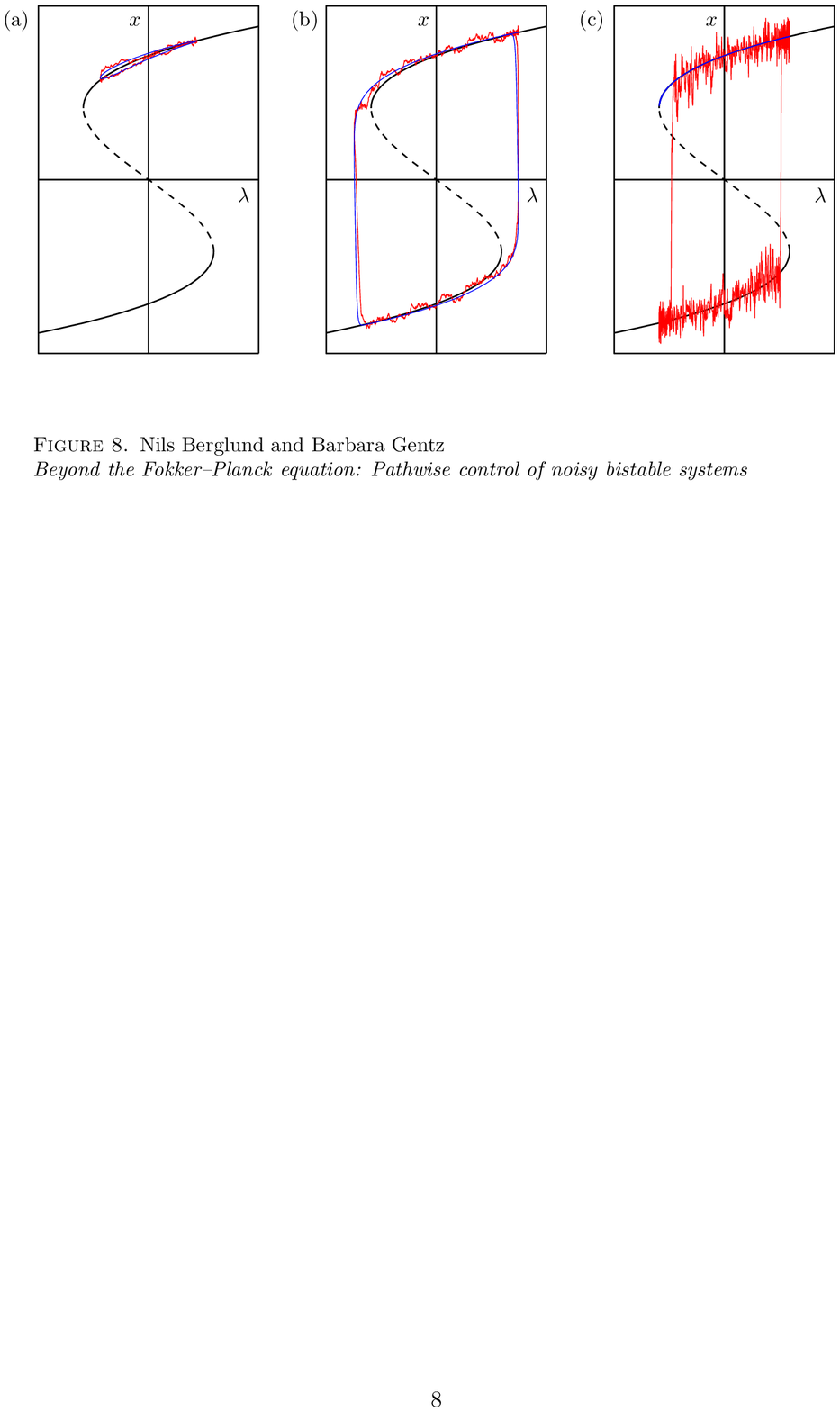,height=60mm,clip=t}}
 \captionspace
 \caption[]
 {Typical random hysteresis \lq\lq cycles\rq\rq\ in the three parameter
 regimes of \figref{fig_hystdiagram}. Deterministic solutions are shown for
 comparison. (a) Case I, small-amplitude regime (here $\eps=0.05$,
 $a_0=-0.1$, $\sigma=0.025$): Paths typically stay close to the
 deterministic cycle, which tracks a potential well. (b) Case II,
 large-amplitude regime (here $\eps=0.02$, $a_0=0.1$, $\sigma=0.05$):
 Typical  paths are close to the deterministic cycle, which switches between
 potential wells. (c) Case III, large-noise regime (here $\eps=0.001$,
 $a_0=0$, $\sigma=0.16$): Paths are likely to cross the potential barrier
 when $\abs{\lambda}$ is of order $\lc-\sigma^{4/3}$.}
\label{fig_hystcycles}
\end{figure}

We thus obtain the existence of three distinct parameter regimes, with
qualitatively different behaviour of typical hysteresis cycles. We
summarize the main results in the following theorem, and give some
additional details afterwards. Many estimates contain logarithmic
dependencies on $a_0$, $\sigma$ and $\eps$. In order not to overburden
notations, we will assume that $\sigma$ and $a_0$ behave like a power of
$\eps$ ($a_0$ may also be a constant), and denote $\abs{\log\eps}$ by
$\ell_\eps$. The regimes are those indicated in \figref{fig_hystdiagram},
but some results are only valid if we exclude a logarithmic layer near the
boundary, for instance Case~II corresponds to $a_0>\gamma_1\eps$ and
$\sigma\leqs\text{{\it const }}\esa0^{1/2}/\ell_\eps$.  

\goodbreak
\begin{theorem}[{{\rm\cite[Theorems~2.3--2.5]{BG3}}}] \hfill
\label{thm_han}
\begin{itemiz}
\item	{\bf Case I:} (Small-amplitude regime)

The distribution of the random area $\cA(\eps,\sigma)$ is concentrated near
the deterministic value $\cA(\eps,0)\asymp A\eps$. There are two subcases to
consider:
\begin{itemiz}
\item	
In Case Ia, $\cA(\eps,\sigma)$ can be written as the
sum of a Gaussian random variable with variance of order
$\sigma^2\eps$, centred at $\cA(\eps,0)$, and a random
remainder. The remainder has expectation and standard deviation
of order $\sigma^2\ell_\eps$ at most.
\item	In Case Ib, the distribution of $\cA(\eps,\sigma)$ is more spread
out. Expectation and standard deviation of $\cA(\eps,\sigma)-\cA(\eps,0)$
are at most of order $\sigma^2\ell_\eps$, which may exceed
$\cA(\eps,0)$. 
\end{itemiz}

\item	{\bf Case II:} (Large-amplitude regime)

The distribution of $\cA(\eps,\sigma)$ is concentrated near the
deterministic value $\cA(\eps,0)$ which satisfies \eqref{hds2}. 
\begin{itemiz}
\item	
In Case IIa, $\cA(\eps,\sigma)$ can be written as the
sum of a Gaussian random variable with variance of order
$\sigma^2\esa0^{1/3}$, centred at $\cA(\eps,0)$, and a random
remainder. The remainder has expectation and standard deviation
of order $\sigma^2\ell_\eps\esa0^{-2/3}$ at most.
\item	In Case IIb, we can only show that the distribution of
$\cA(\eps,\sigma)$ is concentrated in an interval of width
$\esa0^{2/3}\ell_\eps$ around $\cA(\eps,0)$. 
\end{itemiz}
\item	{\bf Case III:} (Large-noise regime)

The distribution of $\cA(\eps,\sigma)$ is concentrated near a
(deterministic) reference area $\hat\cA$ satisfying
$\hat\cA-\cA_0\asymp-\sigma^{4/3}$. The standard deviation of
$\cA(\eps,\sigma)$ is at most of order
$\sigma^{4/3}\smash{\ell_\eps^{2/3}}$, and its expectation belongs to an
interval
\begin{equation}
\label{han4}
\bigbrak{\hat\cA - \Order{\sigma^{4/3}\ell_\eps^{2/3}}, 
\hat\cA + \Order{\sigma^2\ell_\eps^2} + \Order{\eps\ell_\eps}}.
\end{equation}
In the case where $a_0\geqs \eps$ and $\sigma\leqs a_0^{3/4}$, the term
$\Order{\eps\ell_\eps}$ has to be replaced by
$\Order{\eps\sqrt{\abs{a_0}}\mskip1.5mu\ell_\eps/\sigma^{2/3}}$. In
both cases, the distribution decays faster to the right of $\hat\cA$
than to the left.  
\end{itemiz}
\end{theorem}

In Regimes I and II, the main effect of additive noise is to broaden the
distribution of the area, which remains concentrated, however, around the
corresponding deterministic value. In Regime III, on the other hand, the
hysteresis area obeys a completely new scaling law, which is determined by
the noise intensity rather than by frequency and amplitude of the driving
field. 

The Gaussian behaviour of $\cA$ in Cases~Ia and IIa is obtained in the
following way. The deviation $y_t=x_t-\xdet_t$ from the deterministic
solution satisfies an equation of the form \eqref{swn2b} with $g\equiv1$,
whose solution obeys the integral equation
\begin{equation}
\label{han5a}
y_t = \frac\sigma{\sqrt\eps} \int_{t_0}^t \e^{\alpha(t,s)/\eps}\6W_s +
\frac1\eps \int_{t_0}^t \e^{\alpha(t,s)/\eps} b(y_s,s) \6s,
\end{equation}
where $b(y,s)=\Order{y^2}$.  For small values of $y_t$, the first term
dominates the second one. Its contribution to $\cA(\eps,\sigma)-\cA(\eps,0)
= -\int y_u\lambda'(u)\6u$ can be written as 
\begin{equation}
\label{han5}
\frac{\sigma}{\sqrt\eps} \int_{t_0}^{t_0+1} \gamma(t_0+1,s)\6W_s, 
\quad\quad
\text{where}
\quad\quad
\gamma(t,s) \defby -\int_s^t \e^{\alpha(u,s)/\eps}\lambda'(u)\6u. 
\end{equation}
The variance of this term is given by 
\begin{equation}
\label{han6}
\sigma^2\eps \Gamma(t_0+1,t_0),
\qquad\qquad
\text{where}
\qquad\qquad
\Gamma(t,t_0) \defby \frac1{\eps^2} \int_{t_0}^t \gamma(t,s)^2 \6s. 
\end{equation}
The integral $\Gamma(t_0+1,t_0)$ depends only on properties of the
deterministic solution $\xdet_t$ via the curvature $a(t)$. The auxiliary
function $\gamma(t,s)$ can be evaluated by partial integration, its
leading term behaving like $-\eps\lambda'(s)/\abs{a(s)}$. 

In Case~I, $\Gamma(t_0+1,t_0)$ is of order $1$, and thus the contribution
of the linear term to the variance of the area is of order $\sigma^2\eps$.
In Case~Ia, one can show that the Gaussian term dominates the distribution
of $\cA(\eps,\sigma)$ near $\cA(\eps,0)$, in the sense that 
\begin{equation}
\label{han7}
\bigprob{\abs{\cA(\eps,\sigma)-\cA(\eps,0)}\geqs H} \leqs 
\frac C\eps \e^{-\kappa H^2/(\sigma^2\eps)}
\end{equation}
holds for some constants $C,\kappa>0$, and for all $H$ smaller than a
constant times $\sqrt\eps(\abs{a_0}\vee\eps)^{4/3}$ if
$\abs{a_0}\leqs\eps^{2/3}/\smash{\ell_\eps^{4/3}}$, and all $H$ smaller
than  $\eps/\ell_\eps$ if
$\abs{a_0}\geqs\eps^{2/3}/\smash{\ell_\eps^{4/3}}$. Note that the upper
bound \eqref{han7} is exponentially small for the maximal value of $H$,
except on the upper boundary of Region~Ia. 

In Case~Ib, the Gaussian term no longer dominates, but one can still show
that 
\begin{equation}
\label{han8}
\bigprob{\abs{\cA(\eps,\sigma)-\cA(\eps,0)}\geqs H} \leqs 
\frac C\eps \e^{-\kappa H/(\sigma^2\ell_\eps)}
\end{equation}
up to $H = \text{{\it const }}\abs{a_0}^{3/2}\ell_\eps$. Again, \eqref{han8}
is exponentially small except on the upper boundary of Region~Ib. 

\begin{figure}
 \centerline{\psfig{figure=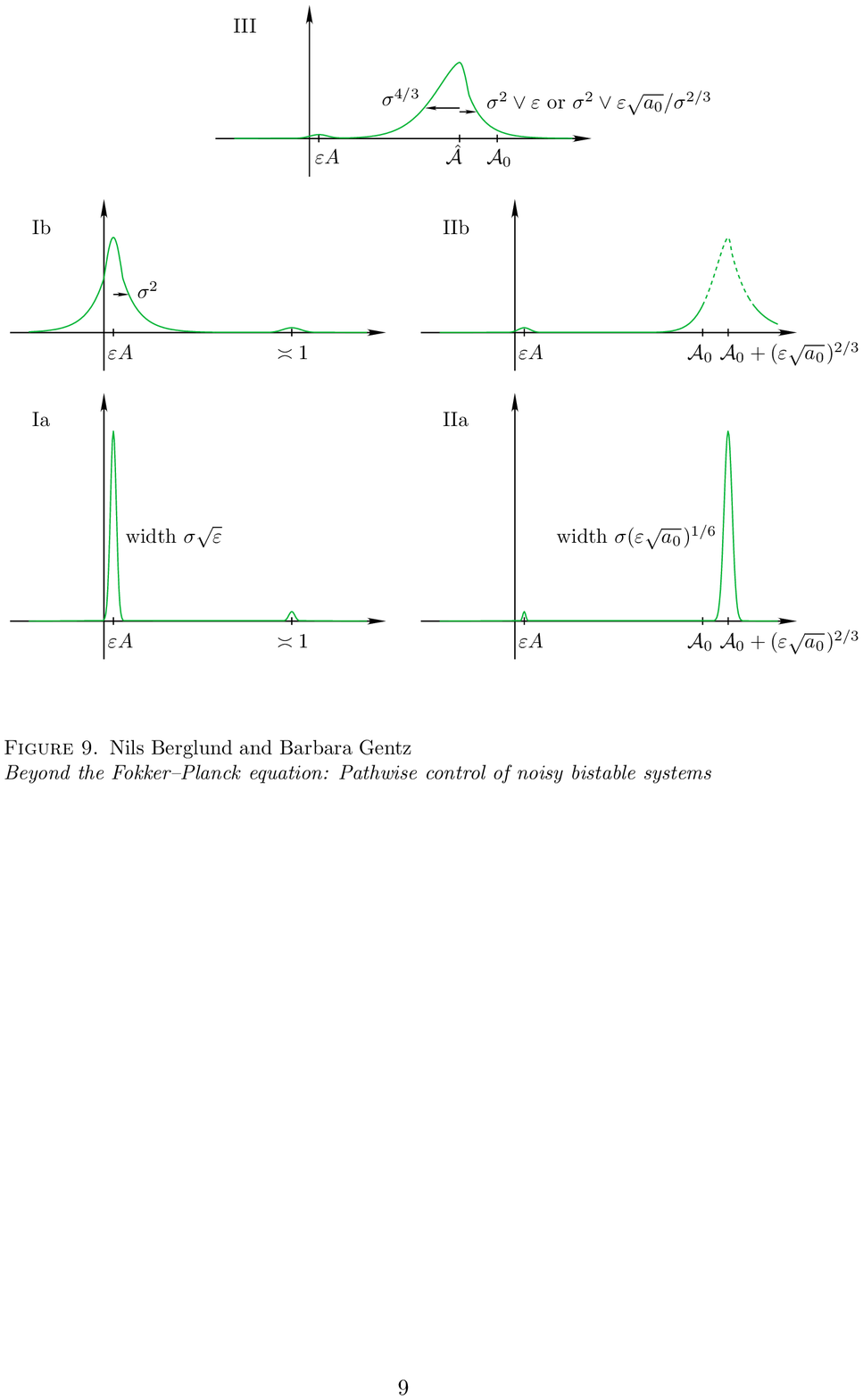,width=150mm,clip=t}}
 \captionspace
 \caption[]
 {Sketches of the distribution of the hysteresis area $\cA(\eps,\sigma)$ in
 the different parameter regimes. Regime~Ia: The area is concentrated near
 the deterministic area $\cA(\eps,0)\asymp A\eps$. There is a small
 probability to observe areas of order $1$. Regime~Ib: The distribution is
 more spread out. Regime~IIa: The area is concentrated near the
 deterministic area $\cA(\eps,0)$ of order $\cA_0+\esa0^{2/3}$. 
 Regime~IIb: The area is concentrated in an interval of width $\esa0^{2/3}$
 around $\cA_0+\esa0^{2/3}$, but we do not control the distribution in this
 interval. The broken curve shows an extrapolation of the estimates outside
 this interval. Regime~III: The area is concentrated near $\hat\cA$, which
 is of order $\cA_0-\sigma^{4/3}$. The distribution decays faster to the
 right than to the left.}
\label{fig_hystdistribution}
\end{figure}

Estimates \eqref{han7} and \eqref{han8} control the tails of the
distribution of $\cA(\eps,\sigma)$ in a \nbh\ of $\cA(\eps,0)$. The quartic
growth of the potential $V(x,t)$ for large $\abs{x}$ implies, on the other
hand, that 
\begin{equation}
\label{han9}
\bigprob{\abs{\cA(\eps,\sigma)-\cA(\eps,0)}\geqs H} \leqs 
\frac C\eps \e^{-\kappa H^4/\sigma^2}
\end{equation}
for all $H$ larger than some constant (of order $1$). In fact, this
estimate holds in {\it all\/} parameter regimes, since it does not depend on
the details of the potential near $x=0$. This still leaves a gap between
the domains of validity of \eqref{han7} and \eqref{han8}, and of
\eqref{han9}, which is due to the existence of a second potential well. In
fact, the distribution of the hysteresis area will not be unimodal. Sample
paths are unlikely to cross the potential barrier, but if they do so, then
most probably near the instants of minimal barrier height, in which case
they enclose an area of order $1$. Hence the density of $\cA(\eps,\sigma)$
will have a large peak near $\cA(\eps,0)$, and a small peak near areas of
order $1$ (more precisely, near $\int_{-A}^A
(X^\star_+(\lambda)-X^\star_-(\lambda))\6\lambda$),
see~\figref{fig_hystdistribution}. 

In Case~IIa, the distribution of $\cA(\eps,\sigma)$ near $\cA(\eps,0)$ is
again dominated by a Gaussian, stemming from the linearization of the SDE
around $\xdet_t$. The integral $\Gamma(t_0+1,t_0)$ in \eqref{han6} is found
to behave like $\eps^{-2/3}\smash{a_0^{1/6}}$, leading to a variance of
order $\sigma^2\esa0^{1/3}$ and to the bound 
\begin{equation}
\label{han10}
\bigprob{\abs{\cA(\eps,\sigma)-\cA(\eps,0)}\geqs H} \leqs 
\frac C\eps \e^{-\kappa H^2/(\sigma^2\esa0^{1/3})},
\end{equation}
valid for $H$ smaller than a constant times $\eps\sqrt{a_0}$. 

Unfortunately, this estimate cannot be extended to Case~IIb. The reason is
that during the jump of $\xdet_t$ to the left-hand potential well, a zone of
instability is crossed where paths are strongly dispersed. The maximal
spreading of paths is of order $\sigma\esa0^{-5/6}$, which is too large, in
this regime, to allow for a precise control of the effect of nonlinear
terms. This does not exclude the possibility that the bound \eqref{han10}
remains valid for larger $\sigma$. 

However, the value $\lambda^0$ of the magnetic field when $x_t$ changes sign
can be described in all of Regime~II. One can show that 
\begin{align}
\label{han11}
\bigprob{\abs{\lambda^0}<\lc-L} &\leqs 
\frac C\eps \e^{-\kappa(\abs{L}^{3/2}\vee\eps\sqrt{a_0}\mskip1.5mu)/\sigma^2}
&&\text{for $-L_1\esa0^{2/3}\leqs L\leqs L_0/\ell_\eps$} \\
\label{han12}
\bigprob{\abs{\lambda^0}>\lc+L} &\leqs 
3 \e^{-\kappa L/(\sigma^2\esa0^{3/2}\ell_\eps)}
&&\text{for $L\geqs L_2\esa0^{2/3}$.} 
\end{align}
Note that $\abs{\lambda^0}$ cannot exceed $\lc+a_0$. These bounds mean that
the distribution of $\lambda^0$ is concentrated around the deterministic
value of order $\lc+\esa0^{2/3}$, and decays faster to the right than to the
left. They can be used to show that in Case~IIb, 
\begin{align}
\label{han13}
\bigprob{\cA(\eps,\sigma)-\cA(\eps,0)\leqs -H} &\leqs 
\frac C\eps \e^{-\kappa H^{3/2}/\sigma^2},\\
\label{han14}
\bigprob{\cA(\eps,\sigma)-\cA(\eps,0)\geqs +H} &\leqs 
\frac C\eps \e^{-\kappa\esa0^{1/3}H/(\sigma^2\ell_\eps)}
\end{align}
for $\esa0^{2/3}\ell_\eps \leqs H \leqs \esa0^{1/3}\ell_\eps$. Thus
the probability that $\cA(\eps,\sigma)-\cA(\eps,0)$ is outside an interval
of size $\esa0^{2/3}\ell_\eps$ is very small. We do not control, however,
what happens inside this interval. 

In Case~III, the large-noise regime, most sample paths are driven over the
potential barrier as soon as the magnetic field reaches a value of order
$\lc-\sigma^{4/3}$. Rather than comparing $\cA(\eps,\sigma)$ to its
deterministic value, we should compare it to a reference area $\hat\cA$
given by 
\begin{equation}
\label{han15}
\frac12 \hat\cA = \int_{-1/4}^{t_1} \xdetof{+}_s (-\lambda'(s))\6s  
+ \int_{t_1}^{1/4} \xdetof{-}_s (-\lambda'(s))\6s,
\end{equation}
where $\xdetof{\pm}_s$ are solutions tracking respectively the right and
left potential well, and $t_1\asymp-\sigma^{2/3}$ is the typical jump time.
Checking that $\hat\cA-\cA_0$ scales like $-\sigma^{4/3}$ is
straightforward. 

The probability of deviations of $\cA(\eps,\sigma)$ from $\hat\cA$ can be
estimated by bounding separately the integrals between $-1/4$ and $t_1$ and
between $t_1$ and $1/4$. The results differ slightly in two regimes. 
If $a_0\leqs\eps$ or $\smash{\sigma>a_0^{3/4}}$, then 
\begin{align}
\label{han16}
\bigprob{\cA(\eps,\sigma)-\hat\cA\leqs -H} &\leqs 
\frac C\eps \e^{-\kappa H^{3/2}/\sigma^2} 
+ \frac32 \e^{-\kappa\sigma^{4/3}/(\eps\ell_\eps)}\\
\label{han17}
\bigprob{\cA(\eps,\sigma)-\hat\cA\geqs +H} &\leqs 
\frac C\eps \e^{-\kappa H/(\sigma^2\ell_\eps)} 
+ \frac32 \e^{-\kappa H/(\eps\ell_\eps)}
\end{align}
holds for some $C,\kappa>0$ and all $H$ up to a constant times
$\sigma^{2/3}\ell_\eps$. If $a_0\geqs\eps$ and $\sigma\leqs a_0^{3/4}$, two
exponents are modified:
\begin{align}
\label{han18}
\bigprob{\cA(\eps,\sigma)-\hat\cA\leqs -H} &\leqs 
\frac C\eps \e^{-\kappa H^{3/2}/\sigma^2} 
+ \frac32 \e^{-\kappa\sigma^2/(\eps\sqrt{a_0}\mskip1.5mu\ell_\eps)}\\
\label{han19}
\bigprob{\cA(\eps,\sigma)-\hat\cA\geqs +H} &\leqs 
\frac C\eps \e^{-\kappa H/(\sigma^2\ell_\eps)} 
+ \frac32 \e^{-\kappa \sigma^{2/3}H/(\eps\sqrt{a_0}\mskip1.5mu\ell_\eps)}.
\end{align}
Again, the distribution decays faster to the right of $\hat\cA$ than to the
left, guaranteeing that $\cA(\eps,\sigma)$ is likely to be smaller than the
static hysteresis area $\cA_0$. The second term on the right-hand sides of
\eqref{han16} and \eqref{han18} does not depend on $H$: It bounds the
probability that the paths do not cross the potential barrier, and enclose
an area close to zero. The situation is opposite to Case~I: The
distribution of the hysteresis area has a large peak near $\hat\cA$ and a
small peak near $0$. 

The qualitative behaviour of the distribution of $\cA(\eps,\sigma)$ is
sketched in \figref{fig_hystdistribution}. When crossing the boundary
between Regime~I and Regime~III, the probability of paths crossing the
potential barrier increases, so that the peak near $\cA(\eps,0)$ shrinks
while the peak near $\cA_0$ grows. When approaching the transition line
between Regimes~III and II, the distribution of the area $\cA(\eps,\sigma)$
becomes more spread out and more symmetric, before concentrating again
around the large-amplitude deterministic value of $\cA_0 +
\Order{\esa0^{2/3}}$. 


\section{Bifurcation delay}
\label{sec_bd}

In the previous section, we had to deal in particular with the slow passage
through a saddle--node bifurcation. This section is devoted to the slow
passage through a (symmetric) pitchfork bifurcation.

We consider again the Ginzburg--Landau potential \eqref{sr1}, but this time
with $\lambda\equiv0$, and a parameter $\mu(t)$ increasing monotonously
through zero. As $\mu$ changes from negative to positive, the potential
transforms from a single-well to a double-well potential, a scenario known
as \defwd{spontaneous symmetry breaking}. In fact, the symmetry of the
potential is not broken, but the symmetry of the state may be. Solutions
tracking initially the potential well at $x=0$ will choose between
one of the new potential wells, but which one of the wells is chosen, and
at what time, depends strongly on the noise present in the system. 


\subsection{Dynamic pitchfork bifurcation}
\label{sec_bdd}

In the deterministic case $\sigma=0$, the equation of motion reads
\begin{equation}
\label{bdd1}
\eps\dtot{x_t}t = \mu(t) x_t - x_t^3.
\end{equation}
Its solution $\xdet_t$ with initial condition $\xdet_{t_0}=x_0>0$ can be
written in the form 
\begin{equation}
\label{bdd2}
\xdet_t = c(x_0,t) \e^{\alpha(t,t_0)/\eps}, 
\qquad\qquad
\alpha(t,t_0) = \int_{t_0}^t \mu(s)\6s,
\end{equation}
where the function $c(x_0,t)$ is found by substitution into \eqref{bdd1}.
Its exact expression is of no importance here, it is sufficient to know that
$0<c(x_0,t)\leqs x_0$ for all $t$. 

Assume that $\mu(t)$ is negative for $t<0$ and positive for $t>0$. If
we start at a time $t_0<0$, the solution \eqref{bdd2} will be
attracted exponentially fast by the stable origin. The function
$\alpha(t,t_0)$ is negative and decreasing for $t_0<t<0$, which
implies in particular that $\xdet_0$ is exponentially small. For
$t>0$, the function $\alpha(t,t_0)$ is increasing, but it remains
negative for some time. As a consequence, $\xdet_t$ remains close to
the saddle up to the first time $t=\Pi(t_0)$ for which $\alpha(t,t_0)$
reaches $0$ again (if such a time exists). Shortly after time
$\Pi(t_0)$, the solution will jump to the potential well at
$+\sqrt{\mu(t)}$, unless $x_0$ is exponentially small. $\Pi(t_0)$ is
called \defwd{bifurcation delay}, and depends only on $\mu$ and
$t_0$. For instance, if $\mu(t)=t$, then $\alpha(t,t_0)=\frac12
(t^2-t_0^2)$ and $\Pi(t_0)=\abs{t_0}$. 

The existence of a bifurcation delay may have undesired consequences. Assume
for instance that we want to determine the bifurcation diagram of
$\dot x=\mu x-x^3$ experimentally. Instead of measuring the asymptotic value
of $x_t$ for many different values of $\mu$, which is time-consuming
(especially near $\mu=0$ where $x_t$ decays only like $1/\sqrt t$), one may
be tempted to vary $\mu$ slowly during the experiment. This, however, will
fail to reveal part of the stable equilibrium branches because of the
bifurcation delay. A similar phenomenon exists for the Hopf bifurcation
\cite{Neishtadt1,Neishtadt2}. 

The delay is due to the fact that $x_t$ approaches the origin
exponentially closely. Noise of sufficient intensity will help driving the
particle away from the saddle, and should therefore reduce the bifurcation
delay. The obvious question is thus: How does the delay depend on the noise
intensity?

In the case of a quadratic potential $V(x,t) = -\frac12\mu(t)x^2$, this
question has been investigated  and compared with experiments, with the
result that the delay behaves like $\smash{\sqrt{\abs{\log\sigma}}}$
\cite{TM,SMC,SHA}. Similar results were obtained in \cite{JL}.  Our
techniques allow us to derive rigorous bounds for the nonlinear equation. 


\subsection{Effect of additive noise}
\label{sec_bdn}

We consider the nonlinear SDE
\begin{equation}
\label{bdn1}
\6x_t = \frac1\eps \bigbrak{\mu(t)x_t - x_t^3}\6t +
\frac\sigma{\sqrt\eps}\6W_t.
\end{equation}
The results presented here are a particular case of those obtained in
\cite{BG1}, which apply to more general nonlinearities.  We assume that
$\mu(t)=t+\Order{t^2}$ is monotonously increasing on an interval $[t_0,T]$
or $[t_0,\infty)$, $t_0<0$. We denote by $\xdet_t$ and $x_t$ the solutions
of the deterministic and stochastic equations with given initial condition
$x_0>0$. 

We already know that $\xdet_t$ decreases exponentially fast for $t<0$.
Results from Section~\ref{sec_sw} show that paths are concentrated in a
\nbh\ of order $\sigma$ of $\xdet_t$ on any time interval $[t_0,t_1]$
bounded away from zero, so that we only need to worry about what happens
after time $t_1$, when $\xdet$ is already exponentially small. 

The dispersion of paths will be controlled by the variance-like function 
\begin{equation}
\label{bdn2}
\bv(t) = \bv_0 \e^{2\alpha(t,t_1)/\eps} + \frac{\sigma^2}\eps \int_{t_1}^t
\e^{2\alpha(t,s)/\eps} \6s,
\end{equation} 
where $\bv_0$ is a positive constant. One can show that this function grows
like $\sigma^2/\abs{\mu(t)}$ for $t\leqs-\sqrt\eps$, and remains of order
$\sigma^2/\sqrt\eps$ up to time $\sqrt\eps$. Only after time $\sqrt\eps$,
$\bv(t)$ grows exponentially fast. In analogy with \eqref{swl15}, we define
a strip
\begin{equation}
\label{bdn3}
\cB(h) = \bigsetsuch{(x,t)}{t_1\leqs t\leqs\sqrt\eps,
\abs{x-\xdet_t}<h\sqrt{\bv(t)}}.
\end{equation}
In order to describe the behaviour for $t\geqs\sqrt\eps$, we further
introduce the domain 
\begin{equation}
\label{bdn4}
\cD(\varrho) = \bigsetsuch{(x,t)}{\sqrt\eps\leqs t\leqs T,
\abs{x}\leqs\sqrt{(1-\varrho)\mu(t)}},
\end{equation}
where $\varrho$ is a parameter in $[0,2/3)$. Note that $\cD(2/3)$ contains
those points in space--time where the potential is concave, while $\cD(0)$
contains the points located between the bottoms of the wells. 

\begin{figure}
 \centerline{\psfig{figure=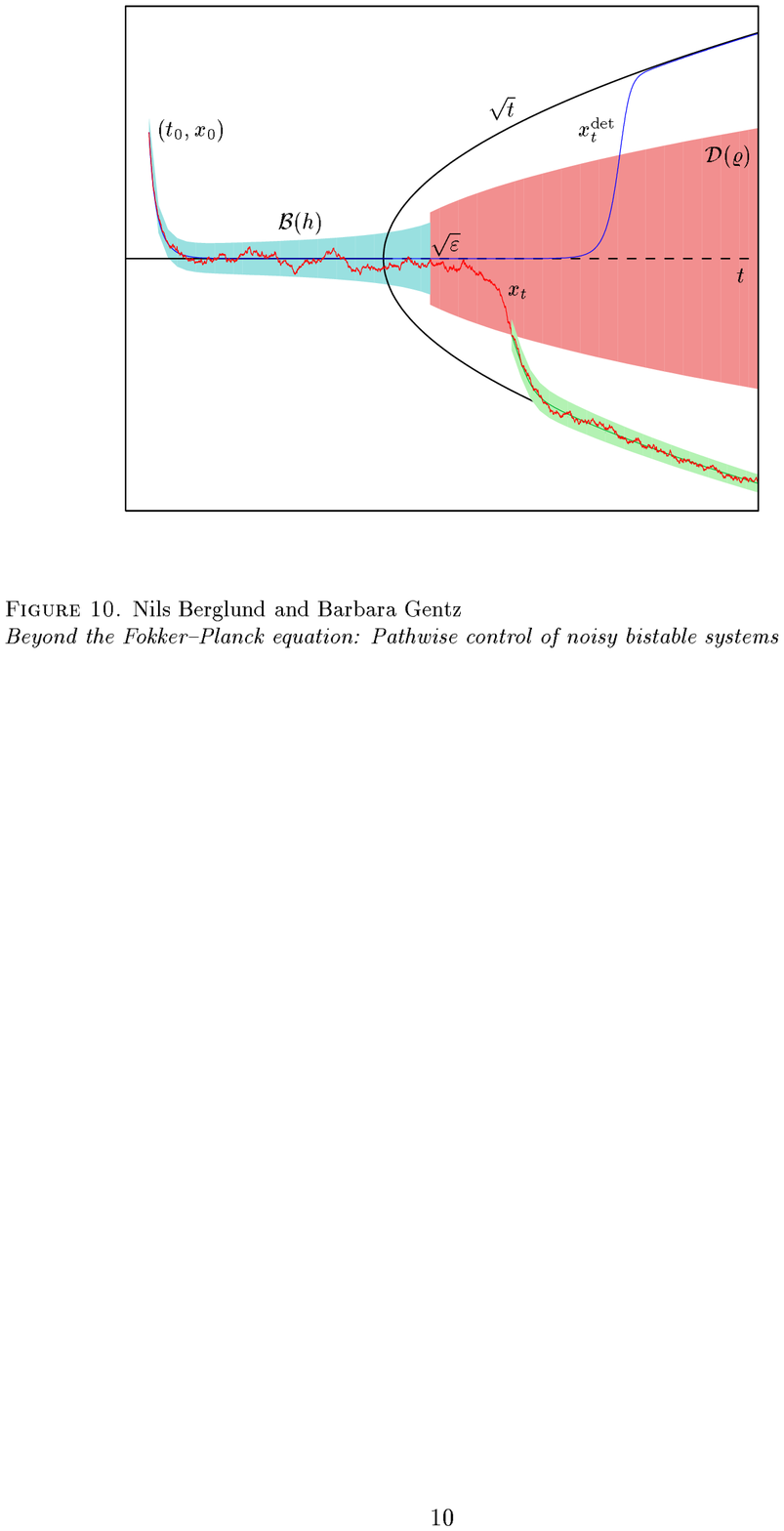,height=80mm,clip=t}}
 \captionspace
 \caption[]
 {A sample path of the SDE~\eqref{bdn1} with $\mu(t)=t$, for $\eps=0.01$ and
 $\sigma=0.015$. A deterministic solution is shown for comparison. Up to
 time $\sqrt\eps$, the path remains in the set $\cB(h)$ centred at
 $\xdet_t$, shown here for $h=3$. It then leaves the set $\cD(\varrho)$
 (here $\varrho=2/3$) after a time of order $\sqrt{\eps\abs{\log\sigma}}$,
 after which it remains in a \nbh\ of the deterministic solution starting
 at the same time on the boundary of $\cD(\varrho)$.}
\label{fig_delay}
\end{figure}

\goodbreak
\begin{theorem}[{{\rm\cite[Theorems~2.10--2.12]{BG1}}}]\hfill
\label{thm_bdn}
\begin{itemiz}
\item	There is a constant $h_0>0$ such that for all $h\leqs
h_0\sqrt\eps/\sigma$, the first-exit time $\tau_{\cB(h)}$ of $x_t$ from
$\cB(h)$ satisfies
\begin{equation}
\label{bdn5}
\bigprobin{t_1,x_{t_1}}{\tau_{\cB(h)}<\sqrt\eps} 
\leqs C_\eps \e^{-\kappa h^2},
\end{equation}
where
\begin{equation}
\label{bdn6}
C_\eps = \frac{\abs{\alpha(\sqrt\eps,t_1)}+\Order{\eps}}{\eps^2} 
\qquad\qquad
\text{and}
\qquad\qquad
\kappa = \frac12 - \Order{\sqrt\eps\mskip1.5mu} 
- \BigOrder{\frac{\sigma^2 h^2}\eps}.
\end{equation}

\item	Assume that $\sigma\abs{\log\sigma}^{3/2} = \Order{\sqrt\eps}$.
Then for any $\varrho\in(0,2/3)$, the first-exit time $\tau_{\cD(\varrho)}$
of $x_t$ from $\cD(\varrho)$ satisfies
\begin{equation}
\label{bdn7}
\bigprobin{\sqrt\eps,x_{\sqrt\eps}}{\tau_{\cD(\varrho)}\geqs t} 
\leqs C(t,\eps) \frac{\abs{\log\sigma}}{\sigma} \frac{\e^{-\varrho
\alpha(t,\sqrt\eps\mskip1.5mu)/\eps}}
{\sqrt{1-\e^{-2\varrho\alpha(t,\sqrt\eps\mskip1.5mu)/\eps}}},
\end{equation}
where
\begin{equation}
\label{bdn8}
C(t,\eps) = \text{{\it const }} \mu(t)
\Bigpar{1+\frac{\alpha(t,\sqrt\eps\mskip1.5mu)}\eps}. 
\end{equation}

\item	Assume $x_t$ leaves $\cD(\varrho)$ (with $1/2<\varrho<2/3$)
through its upper (lower) boundary. Let $\smash{\xdetof{\tau}_t}$ be
the deterministic solution starting at time $\tau=\tau_{\cD(\varrho)}$
on the upper (lower) boundary of $\cD(\varrho)$. Then
$\smash{\xdetof{\tau}_t}$ approaches the equilibrium branch at
$\smash{\sqrt{\mu(t)}}$ (resp., $-\smash{\sqrt{\mu(t)}}$) like
$\eps/\mu(t)^{3/2} + \smash{\sqrt{\mu(\tau)}}
\e^{-\eta\alpha(t,\tau)/\eps}$, where $\eta=2-3\varrho$. Moreover,
$x_t$ is likely to stay in a strip centred at $\xdetof{\tau}_t$, with
width of order $\sigma/\smash{\sqrt{\mu(t)}}$, at least up to times of
order $1$. 
\end{itemiz}
\end{theorem}

The bound \eqref{bdn5}, which is proved in a similar way as
Theorem~\ref{thm_swn}, shows that paths are unlikely to leave the strip
$\cB(h)$ if $1\ll h\leqs h_0\sqrt\eps/\sigma$. If $\sigma$ is smaller than
$\sqrt\eps$, paths remain concentrated in a \nbh\ of the origin up to time
$\sqrt\eps$, with a typical spreading growing like $\sigma/\sqrt{\abs{\mu(t)}}$
for $t\leqs-\sqrt\eps$, and remaining of order $\sigma/\eps^{1/4}$ for
$\abs{t}\leqs\sqrt\eps$. This is again a dynamical effect: Although there is
a saddle at the origin for positive times, its curvature is so small that
paths do not have time to escape before $t=\sqrt\eps$, see~\figref{fig_delay}.

Relation~\eqref{bdn7} yields an upper bound on the typical time needed to
leave $\cD(\varrho)$, and thus enter a region where the potential is
convex. Since $\alpha(t,\sqrt\eps)$ grows like $\frac12 t^2$ for small $t$,
the probability not to leave $\cD(\varrho)$ before time $t$ becomes small as
soon as
\begin{equation}
\label{bdn9}
t \gg \sqrt{\frac2\varrho \eps\abs{\log\sigma}}. 
\end{equation}
The last part of the theorem implies that another time span of the same
order is needed for paths to concentrate again, around an adiabatic 
solution tracking the bottom of the well (at a distance of order
$\eps/\mu(t)^{3/2}$). One can thus say that the typical bifurcation
delay time of the dynamical pitchfork bifurcation with noise is of order
$\sqrt{\eps\abs{\log\sigma}}$. 

\begin{figure}
 \centerline{\psfig{figure=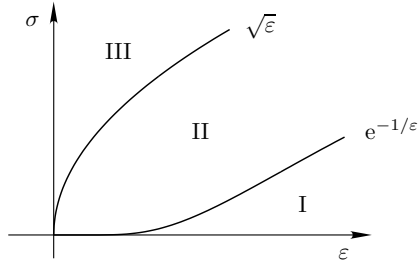,width=55mm,clip=t}}
 \captionspace
 \caption[]
 {Depending on the value of noise intensity $\sigma$ and drift velocity
 $\eps$, the random bifurcation delay has one of three qualitatively
 different behaviours. If $\sigma$ is exponentially small in $1/\eps$, the
 typical delay is macroscopic. For $\sigma$ larger than $\sqrt\eps$, there
 is no such delay, and paths make large excursions near the
 bifurcation point. In  the intermediate regime, the typical delay is
 of order $\sqrt{\eps\abs{\log\sigma}}$.}
\label{fig_delayregimes}
\end{figure}

\goodbreak
As a consequence, we can distinguish three parameter regimes
(\figref{fig_delayregimes}):
\begin{enum}
\item[I.] {\it Exponentially small noise:\/} $\sigma\leqs\e^{-K/\eps}$
for some $K>0$. 

At time $\sqrt\eps$, the spreading of paths is still exponentially small.
In fact, one can extend Relation~\eqref{bdn5} to all times for which
$\alpha(t,\sqrt\eps\mskip1.5mu) < K$. If $K$ is larger than
$\alpha(\Pi(t_0),0)$, where $\Pi(t_0)$ is the deterministic bifurcation
delay, most paths will track the deterministic solution and follow it into
the right-hand potential well. 

\item[II.] {\it Moderate noise:\/}
$\e^{-1/\eps^p}\leqs\sigma\ll\sqrt\eps$ for some $p>1$. 

The bifurcation delay lies between $\sqrt\eps$ and a constant times
$\sqrt{\eps\abs{\log\sigma}} \leqs \eps^{(1-p)/2}$ with high probability. 
One can thus speak of a \lq\lq microscopic\rq\rq\ bifurcation delay. 

\item[III.] {\it Large noise:\/} $\sigma\geqs\sqrt\eps$.

The spreading of paths grows like $\sigma/\sqrt{\abs{\mu(t)}}$ at least up to
time $-\sigma$. As $t$ approaches the bifurcation time $0$, the bottom of
the potential well becomes so flat that the paths are no longer localized
near the origin and may switch wells several times before eventually
settling for a well. So for large noise intensities, the concept of
bifurcation delay should be replaced by {\it two\/} variables, namely the
first-exit time from a suitably chosen \nbh\ of the saddle and the time
when the potential wells become attractive enough to counteract the
diffusion.
\end{enum}

One can also estimate the probability to reach the right-hand potential well
rather than the left-hand potential well. Loosely speaking, if $x_s$ reaches
$0$ before time $t$, it has equal probability to choose either potential
well. It follows that 
\begin{equation}
\label{bdn10}
\bigprobin{t_0,x_0}{x_t\geqs 0} = 
\frac12 + \frac12 \bigprobin{t_0,x_0}{x_s > 0\;\forall s\in[t_0,t]}.
\end{equation}
One can show that for $t=0$ the second term on the right-hand side is of
order 
\begin{equation}
\label{bdn11}
\frac{x_0\eps^{1/4}}\sigma \e^{-\abs{\alpha(0,t_0)}/\eps},
\end{equation}
and for larger $t$, this term will be even smaller. Thus in case~II, 
paths will choose one potential well or the other with a probability
exponentially close to $1/2$. 

In the particular case where $\mu(t)>\text{{\it const }}t$ for all $t>0$, the
height of the potential barrier grows without bound. This implies that once
$x_t$ has chosen a potential well, its probability {\it ever\/} to cross the
saddle again is of order $\e^{-\text{{\it const }}/\sigma^2}$.

The existence of three parameter regimes has some interesting consequences
on the experimental determination of a bifurcation diagram. Assume we want
to determine the stable equilibrium branches by sweeping the parameter with
speed $\eps$. Regime~II is the most favourable: In Regime~I, part of the
stable branches cannot be seen due to the bifurcation delay, while in
Regime~III, noise will blur the bifurcation diagram. For a given noise
intensity $\sigma$, the sweeping rate $\eps$ should thus satisfy 
\begin{equation}
\label{bdn12}
\sigma^2 \ll \eps \ll \bigl(1/\abs{\log\sigma}\bigr)^{1/p}
\end{equation}
in order to produce a good image of the stable equilibria. As long as 
$\sigma^2\ll1/\abs{\log\sigma}^{1/p}$, increasing artificially the noise level
allows to work with higher sweeping rates, but of course the image
will be more and more blurred.

On the other hand, the relation between noise and delay can be used to
measure the intensity of noise present in the system. If a bifurcation delay
is observed for a sweeping rate $\eps_0$, repeating the experiment with
slower and slower sweeping rates should ultimately suppress the delay. If
this happens for $\eps=\eps_1$, then the noise intensity is of order
$\e^{-\text{{\it const }}/\eps_1}$.


\section{Generalizations}
\label{sec_g}


\subsection{Multidimensional systems}
\label{ssec_gm}

We now return to $n$-dimensional equations such as \eqref{in11}. As in
Section~\ref{sec_sw}, we start by examining the linear equation
\begin{equation}
\label{gm1}
\6y_t = \frac1\eps A(t) y_t \6t + \frac\sigma{\sqrt\eps} G(t)\6W_t, 
\qquad \qquad y_0=0,
\end{equation}
obtained, for instance, by linearizing the equation around a given
deterministic solution (in that case, the matrices $A$ and $G$ may depend on
$\eps$). If the drift term derives from a potential, then $A$ is necessarily
symmetric, but we will not impose such a restriction here. We will assume,
however, that we are in the stable case, that is, all eigenvalues of $A(t)$
have real parts smaller than some constant $-a_0<0$. 

The solution of \eqref{gm1} can be written as 
\begin{equation}
\label{gm2}
y_t = \frac\sigma{\sqrt\eps} \int_0^t U(t,s) G(s) \6W_s,
\end{equation}
where $U(t,s)=U(t)U(s)^{-1}$ is the propagator of the deterministic
equation $\eps\dot y=A(t)y$, and we denote by $U(t)$ its principal
solution, i.\,e., $\eps\dot U(t)=A(t)U(t)$ and $U(0)=\one$. The random
variable $y_t$ has a Gaussian distribution, with zero expectation and
covariance matrix
\begin{equation}
\label{gm3}
\covar\set{y_t} = \frac{\sigma^2}\eps \int_0^t U(t,s) G(s) \transpose{G(s)}
\transpose{U(t,s)} \6s.
\end{equation}
In order to determine the asymptotic behaviour of the covariance matrix, 
we first note that it is a particular solution of the ODE 
\begin{equation}
\label{gm4}
\eps \dtot{}{t} X(t) = A(t) X(t) + X(t) \transpose{A(t)} + \sigma^2 G(t)
\transpose{G(t)}. 
\end{equation}
The general solution of \eqref{gm4} can be written as 
\begin{equation}
\label{gm5}
X(t) = \Xbar(t) + U(t) \bigbrak{X(0) - \Xbar(0)} \transpose{U(t)},
\end{equation}
where $\Xbar(t)$ is a particular solution admitting an asymptotic
expansion (c.\,f.~\cite{Wasow}) 
\begin{equation}
\label{gm6}
\Xbar(t) = \Xbar_0(t) + \eps \Xbar_1(t) + \eps^2 \Xbar_2(t) + \dotsb
\end{equation}
The terms of this expansion satisfy the Liapunov equations 
\begin{align}
\label{gm7}
A(t)\Xbar_0(t) + \Xbar_0(t)\transpose{A(t)} &= -\sigma^2 G(t)
\transpose{G(t)} \\ 
\label{gm8}
A(t)\Xbar_n(t) + \Xbar_n(t)\transpose{A(t)} &= \dtot{}{t} \Xbar_{n-1}(t)
\qquad \qquad \forall n\geqs1.
\end{align}
The solution of \eqref{gm7} admits the integral representation (see for
instance~\cite{Bellman})
\begin{equation}
\label{gm9}
\Xbar_0(t) = \sigma^2 \int_0^\infty \e^{A(t)s} G(t)\transpose{G(t)}
\e^{\transpose{A(t)}s} \6s,
\end{equation}
and similarly for the solutions of \eqref{gm8}. Note that $\Xbar_0$
corresponds to the asymptotic covariance matrix of~\eqref{gm2} if
$A$ and $G$ are constant in time

At any fixed time $t$, the distribution of $y_t$ is concentrated in an
ellipsoid of the form $\pscal{y}{\covar\set{y_t}^{-1}y} \leqs
\text{{\it const}}$, provided the (Gaussian) distribution of $y_t$ is
nondegenerate. Assume for the moment that $G(t)\transpose{G(t)}$ is
uniformly positive definite. Then $\covar\set{y_t}^{-1}$ exists and
so does $\Xbar(s)^{-1}$. In addition, we find that
$\sigma^2\Xbar(s)^{-1}$ is bounded in norm. The following result,
which generalizes Proposition~\ref{prop_swl} from the one-dimensional
case, shows that the paths $\set{y_t}_{t\geqs0}$ are  
concentrated in sets $\setsuch{(t,y)}{\pscal{y}{\Xbar(t)^{-1}y} \leqs
\text{{\it const}}}$. The proof will be given in~\cite{BG4}. 

\begin{theorem}
\label{thm_gm}
Assume that $\sigma^2\Xbar(s)^{-1}$ is bounded in norm. Then for all $t>0$,
$h>0$ and any $\kappa\in(0,1/2)$, 
\begin{equation}
\label{gm10}
\Bigprobin{0,0}{\sup_{0\leqs s\leqs t} \pscal{y_s}{\Xbar(s)^{-1} y_s} \geqs
h^2} \leqs C_n(t,\eps) \e^{-\kappa h^2(1-\Order{\eps})},
\end{equation}
where 
\begin{equation}
\label{gm11}
C_n(t,\eps) = \Bigpar{\frac t{\eps^2} + 1}  \Bigpar{\frac1{1-2\kappa}}^n.
\end{equation}
\end{theorem}

The exponential growth of the prefactor as a function of the dimension
implies that the estimate~\eqref{gm10} is only useful for
$h>\sqrt{n}$. This dependence, however,  is to be expected as the tails of
$n$-dimensional (standard) Gaussians only show their typical decay
outside a ball of radius proportional to $\sqrt{n}$.

Under the condition that $\sigma^{-2}\Xbar(s)$ and
$\sigma^2\Xbar(s)^{-1}$  are bounded in norm, one can show 
that a bound similar to~\eqref{gm10} holds if $y_s$ obeys a 
nonlinear perturbation of Equation~\eqref{gm1}, for all $h$ smaller than a
constant times $\sigma^{-1}$. 


\subsection{Coloured noise}
\label{ssec_gc}

White noise has no time-correlations, and is thus appropriate to model the
random influence of a \lq\lq fast\rq\rq\ environment on a \lq\lq slow\rq\rq\
system, if the relaxation time of the fast system is negligible. If the
relaxation time is short but not negligible, one has to use coloured noise
instead of white noise. 

The simplest (still Markovian) model of coloured noise is the
Ornstein--Uhlenbeck process
\begin{equation}
\label{gc1}
Z_t = \sigma \int_0^t \e^{-\gamma(t-s)} \6W_s, 
\end{equation}
whose autocorrelation function $\expec{Z_sZ_t} =
(\sigma^2/2\gamma)\e^{-\gamma (t-s)}[1-\e^{-2\gamma s}]$, $s<t$, decays
exponentially in $t-s$.

Let us examine the influence of such  coloured noise on a slowly
time-dependent, one-dimensional, linear system, described by the SDE
\begin{equation}
\label{gc2}
\6x_t = a(\eps t)x_t \6t + g(\eps t) \6Z_t.
\end{equation}
We assume that $a(\eps t)$ is negative, $g(\eps t)$ is positive, and
that both functions are uniformly bounded away from zero. 
Using the differential representation $\6Z_t=-\gamma Z_t\6t +
\sigma\6W_t$ of the Ornstein--Uhlenbeck process \eqref{gc1}, and
scaling time by a factor $\eps$, we obtain that the random variable
$y_t=(x_t,Z_t)$ obeys a two-dimensional linear SDE of the form 
\eqref{gm1}, with  
\begin{equation}
\label{gc3}
A(t) = 
\begin{pmatrix}
a(t) & - \gamma g(t) \\
0 & -\gamma
\end{pmatrix}
\qquad
\text{and}
\qquad 
G(t) = 
\begin{pmatrix}
g(t) \\ 1
\end{pmatrix}.
\end{equation}
For $\gamma\to0$, we recover the white-noise case, while $\sigma\to0$ 
corresponds to the deterministic limit. 

Theorem~\ref{thm_gm} shows that paths are concentrated in a tube
$\pscal{y_s}{\Xbar(s)^{-1}y_s} \leqs \text{{\it const}}$, with 
covariance matrix $\Xbar(s)$ admitting the asymptotic series \eqref{gm6}. The
leading term $\Xbar_0(s)$ is found from \eqref{gm9} to be 
\begin{equation}
\label{gc4}
\Xbar_0(s) = \sigma^2 
\begin{pmatrix}
\vrule height 11pt depth 16pt width 0pt
\dfrac{g(s)^2}{2(\gamma+\abs{a(s)})} & 
\dfrac{g(s)}{2(\gamma+\abs{a(s)})} \\
\vrule height 16pt depth 11pt width 0pt
\dfrac{g(s)}{2(\gamma+\abs{a(s)})} &
\dfrac1{2\gamma}
\end{pmatrix}.
\end{equation}
In particular, the paths $\set{x_s}_{s\geqs 0}$ are concentrated in a strip of
width $\sigma g(s)/\sqrt{2(\gamma+\abs{a(s)})}$. The effect of coloured
noise of the form \eqref{gc1} is thus to narrow the distribution of the
paths, in the same way as if the curvature were increased by an amount
$\gamma$. 

This result allows us to make some predictions on the effect of noise colour
on stochastic resonance. We have seen in Section~\ref{ssec_srp} that in the
standard case of the asymmetric double-well potential with additive
periodic forcing and white noise, paths switch between potential wells
when $\sigma$ exceeds the threshold $\sigmac=(a_0\vee\eps)^{3/4}$. 
This value was obtained by comparing the typical spreading of paths to
the minimal distance between the deterministic solution and the
saddle. Repeating this argument in the case of coloured noise given
by~\eqref{gc1}, we thus expect the threshold noise intensity to be
determined by
\begin{equation}
\label{gc5}
\sigmac^2 =  (a_0\vee\eps)\bigpar{\gamma \vee (a_0 \vee \eps)^{1/2}}.
\end{equation}
This shows in particular that shorter correlation times of the noise
require larger noise intensities to enable transitions between the wells.

A similar stabilizing effect of noise colour can be expected when sweeping
the bifurcation parameter through a pitchfork bifurcation as discussed in
Section~\ref{sec_bd}. The bifurcation delay will lie between the
macroscopic delay of the deterministic case and the microscopic delay
of the white-noise case. The shorter the correlation time of the
noise, the larger will the delay be. 

\goodbreak
\vfil\eject

\goodbreak

\bigskip\bigskip\noindent
{\small 
Nils Berglund \\ 
{\sc Department of Mathematics, ETH Z\"urich} \\ 
ETH Zentrum, 8092~Z\"urich, Switzerland \\
{\it E-mail address: }{\tt berglund@math.ethz.ch}

\bigskip\noindent
Barbara Gentz \\ 
{\sc Weierstra\ss\ Institute for Applied Analysis and Stochastics} \\
Mohrenstra{\ss}e~39, 10117~Berlin, Germany \\
{\it E-mail address: }{\tt gentz@wias-berlin.de}
}


\end{document}